# A Survey on Privacy of Health Data Lifecycle: A Taxonomy, Review, and Future Directions

Sunanda Bose[a], Dusica Marijan[a]

[a]*Simula Research Laboratory, Oslo, Norway*

**ABSTRACT**

With the increasing breaches and security threats that endanger health data, ensuring patients' privacy is essential. To that end, the research community has proposed a wide variety of privacy-preserving approaches based on cryptography, hashing, or ledger technologies for alleviating health data vulnerability. To establish a comprehensive understanding of health data privacy risks, as well as the benefits and limitations of existing privacy-preserving approaches, we perform a detailed review of existing work and distill 10 distinct privacy concerns occurring in a health data lifecycle. Furthermore, we classify existing approaches based on their applicability to particular privacy concerns occurring at a particular lifecycle stage. Finally, we propose a taxonomy of techniques used for privacy preservation in healthcare and triangulate those techniques with the lifecycle stages and concerns. Our review indicates heavy usage of cryptographical techniques in this domain. However, we have also found that healthcare systems have special requirements that require novel cryptographic techniques and security schemes to address special needs. Therefore, we identify several future research directions to mitigate the security challenges for privacy preservation in health data management.

## 1. Introduction

Health data (HD) refers to health information about a human entity. The definition of HD may vary depending on the legislation. EU legislative act[3] defines it as personal data concerning past, present, or future physical or mental health status that has been collected in the course of registration, or provision of healthcare services. It also includes information derived from the body of a natural person or bodily substances e.g. genetic data or biological samples. According to the GDPR, data concerning health falls under a special category of personal data, also referred to as sensitive data, because it requires additional protection, due to the fact that it can go to the very core of a human being. Unauthorized disclosure of this type of data can cause discrimination and the violation of fundamental rights. This is what makes HD different from other types of personal data, and consequently requires specific protection during data processing, to avoid creating risks to the fundamental rights and freedom. Depending on the mode of creation, HD can be stored in the custody of the patient, healthcare professional, or with an organization for the short or long term. However, HD may include sensitive information about the patient involved, which might put the patient in an uncomfortable situation if accessed by unintended entities. Hence, *ensuring the privacy of patients' sensitive data is an essential requirement in the process of managing HD*. At the same time, there have been incidents of compromised confidentiality of 112 million health records in 2015, threatening the privacy of patients [85]. It has been found that HD from wearable devices can be retrieved from the airwaves [13]. Sensitive medical information is considered a high-value commodity and there is a big data broker industry monetizing the leaked HD from multiple sources. Another study [34] found out 12 apps sent sensitive healthcare information to 76 third parties. The Federal Trade Commission (FTC) has alleged [69] that one Fertility Tracking application has handed users' medical information out to numerous third parties, including Google, Fabric, AppsFlyer, and an analytics firm Flurry. In 2020, 90,000 new health apps were launched in the mobile application market [106]. The expansion of the digital healthcare landscape makes more people vulnerable to privacy risks. In [132] the authors partnered with multi-campus and state-wide hospitals to analyze the traffic coming from all devices in the Network. The study found that there has been the usage of insecure and broken hashes in TLS/SSL usage. Electronic surveillance on accused and convicted criminal offenders in the United States has increased nearly 140% from 2005 to 2015[95][140]. This includes individuals under pre-trial or immigrant detention, house arrest, probation, or parole[7]. Tracking devices include ankle monitors, as well as mobile app-based monitoring solutions purchased from private companies by individuals[7]. Study finds that such Electronic surveillance is not an alternative *to* incarceration, it is an alternative *form* of incarceration[140] such as jail and prison[5].

In order to enforce the privacy of the data in the management of HD, governments of different countries have implemented different legal regulations, such as the Health Insurance Portability and Accountability Act (HIPAA), General Data Protection Regulation (GDPR), and Protection of Personal Information Act (PoPIA). Some of these regulations specifically apply to HD and some apply to general personal data management. In spite of the legal regulations, there are instances of privacy breaches due to the lack of adequate understanding of the agreements mentioned by the regulations.

*Corresponding author

✉ sunanda@simula.no (S. Bose); dusica@simula.no (D. Marijan)

ORCID(s): 0000-0002-6484-9682 (S. Bose); 0000-0001-9345-5431 (D. Marijan)

As a result, although the individuals exchange the information willingly, they may not adequately understand the circumstances and conditions[134]. In [134], this phenomenon is considered Surrendering Information. Furthermore, the legal regulations themselves may contain loopholes, as identified in [12]. For example, data collected by a wearable device is subject to compliance requirements unless the manufacturer is engaged in the transmission of data on behalf of a covered entity, such as Healthcare Institutions (HI) (e.g. Hospital). This implies that, if a patient purchases a wearable device from a retailer and then uses it for sharing body vitals with doctors, then such transactions are not regulated by privacy regulations. Although the HIPAA Privacy Rule includes a set of individually identifiable health information, there are other behavioral and sensor-derived data that can be used for re-identification purposes. Data from wearable devices are often used in judicial proceedings. In such cases, that data can become part of publicly searchable information in some states. The authors in [12] identify ambiguities in the legal status of shared data when business entities change ownership or go bankrupt. The authors mention previous incidents of Radio Shack, who wanted to sell data to pay off its debt, while Apple and AT&T filed a motion to prevent it. The authors also mention the Snowden revelations regarding US intelligence agencies' surveillance of commercial Internet services which endangers the data sovereignty of the parties involved in the transaction of HD.

The privacy issues relating to HD become more concerning with the wide adoption of digitized health care and wearable devices. Digitization makes sensitive health information a copyable construct that is stored and transferred electronically, making it vulnerable to internal and external attacks. Authors in [17] identify several threats to genomic data (as a type of HD) sharing. Although patients' genomic data may be beneficial for precision medicine and personalized treatments, inappropriate use of such data may lead to serious privacy infringement of patients and their blood relatives. In [135] the authors have identified privacy attacks on genomic data based on different settings, such as healthcare, research, direct-to-consumer, and forensic. Different types of attacks discussed in the paper include re-identification, genotype imputation, genotype inference, genotype reconstruction, non-genotypic attribute inference, membership inference, familial search, and kin genotype reconstruction. The paper mentions several incidents of privacy leaks with genomic data. It mentions a long-range familial search, by the US Federal Bureau of Investigation (FBI) to identify the suspect's family members. It also mentions membership inference using summary statistics of genome-wide association studies (GWAS).

To this end, the research community has proposed various technological solutions, ranging from cryptography to hashing and ledger techniques. In this paper, we study the scientific literature regarding the application of different techniques used for ensuring HD privacy during its lifecycle. We propose a taxonomy that classifies existing solutions to ensure HD privacy at different stages of the lifecycle based on the unique challenges occurring at each particular stage. Finally, we identify a set of open research problems derived from the analysis of the solutions and challenges identified in existing research work.

### 1.1. State-of-the-art and Contributions

Along with the proliferation of technological solutions addressing the privacy challenges of HD, researchers have analyzed and synthesized existing knowledge on this topic in a number of survey studies. However, these surveys cover only the partial HD lifecycle, for example, data privacy risks occurring only during data storage and sharing [10] or they present a limited set of solutions for ensuring data privacy. Specifically, the authors in [10] present a survey of privacy-related issues on genomic data. The authors first identify three core categories of problems in this field, privacy-preserving sharing, secure computation and storage, and privacy of query and output. Next, the authors discuss four cryptographic techniques that have been used to address these problems, i.e., homomorphic encryption, garbled circuits, secure hardware, and differential privacy. The authors in [89] investigate the security and privacy (S&P) issues related to the privacy of e-health systems and review technical solutions that have been grouped into four categories, i.e., E-Health data, medical devices, medical network, edge, fog, and cloud. The paper proposes a taxonomy for Security Concerns, Security Requirements, and privacy-related Security Solutions. Security Concerns consist of Unauthorized Access, Data Disclosure, Data Tampering, and Data Forgery. Security Requirements consist of Access Restriction, Confidentiality, Anonymity, Availability, and Accountability. Security Solutions consist of Access Control, Cryptography, Anonymization, Blockchain, Steganography, and Watermarking. The authors present another similar taxonomy for medical devices, medical network, and edge/fog/cloud. The authors in [6] propose a lifecycle of HD consisting of four different stages, such as collection, transformation, modeling, and knowledge creation. Legal regulations of different countries have been discussed in the context of healthcare. The authors discuss three different privacy-preserving methods, i.e., de-identification, hybrid execution, and identity-based anonymization, including k-anonymity, l-diversity, and t-closeness. The authors in [21] present a chronology of medical device security. The paper mostly concentrates on data collection by medical devices. The timeline of threats and vulnerabilities has been categorized into four periods between 1980 and 2006. The authors in [30] address five relevant questions for S&P in the context of healthcare. The questions are, why is healthcare vulnerable and why is it targeted, what are the current threats and consequences, what is the role of the standards and legislation, and how can the health sector move forward? The authors in [57] discuss S&P issues in healthcare while categorizing the healthcare system into multiple generations. Generation 1.0 refers to manual record keeping which is replaced by electronic record keeping in Generation 2.0. Generation 3.0 involves wearable devices and monitoring. Real-time uninterrupted service using electronic healthcare is considered to be generation 4.0. The authors categorize S&P schemes into eight broad categories as follows: processing-based, machine learning (ML)-based, Wearable device-based, IoT-based, telehealthcare-based, policy-based, authentication-based, and network traffic-based. Finally, the authors in [22] discuss S&P issues associated with implanted medical devices e.g. pacemakers, defibrillators, neurostimulators, infusion pumps, and other biosensors. The authors group adverse events based on the type of the device and associate the adverse effects. ISO/IEC 29100:2011(E) standard [29] provides a list of privacy principles that can be followed for assessing the state of privacy preservation in a system. In [40] the authors have used these principles to asses the technical works related to privacy preservation in the Internet of Medical Things (IoMT). The authors considered four stages in the lifecycle of data in IoMT, collection, transmission, storage, and process. The technical works have also been categorized by the architecture of the system as centralized, decentralized, hybrid, or third-party-based. In [91] privacy issues originating from Electronic monitoring mobile apps have been discussed. The study reveals that most apps collect more data (e.g. activity recognition, audio) than the ankle monitors. Moreover, some of these apps request more permission than others. The most privileged app (Sprokit) requested 14 dangerous (according to Android API documentation) permissions and had the most third-party library integrations, while the least privileged app (Uptrust) requests only 8. Moreover, the authors find out that it is possible to identify whether one individual is under EM surveillance or not by a passive observer on the same Wi-Fi network or an entity such as an ISP. Such identification raises concerns about discrimination in society. In [4] a survey of techniques applied in ensuring S&P of HD is presented. The authors categorize the data security techniques applied to healthcare into four broad categories, including Blockchain, cryptography, Biometrics, and Data Hiding techniques. In [71] three major concerns about HD are discussed: Confidentiality, Integrity, and Availability. Whereas in [86] [151] a set of five S&P goals are considered in the context of ensuring the S&P of HD, including Authentication, Confidentiality, Integrity, Non-repudiation, and Availability. In [151] the authors first describe different challenges of privacy preservation in Healthcare and then discuss the existing solutions and practices. The authors in [151] also present case studies of three different scenarios and the relevant privacy challenges. On the other hand, the paper [86] takes an attack-based perspective on the state of the art. It presents a mapping between the attacks and the defense mechanisms while discussing the literature. In [87] a survey on the applications of Federated Learning (FL) in the healthcare domain is presented. The authors categorize Federated learning into three different categories: Horizontal, Vertical, and Federated Transfer Learning. The paper describes three different focuses of FL-related works: managing the resources related to FL, securing the learning process, ensuring privacy while learning, an incentive mechanism, and the personalization of learned models. The security attacks on health care is reviewed in [155]. The authors categorized the attacks into three primary domains, Body Area, Communication and Infrastructure, and Service domain. The attacks in the first category are further categorized as Masquerade attacks through fake identity, Attacks on wearable and implantable medical devices through eavesdropping, impersonation etc., attacks on Body-coupled communications through device cloning, Accountability, and revocability attack by means of abuse of access privileges, Data injection attack with the intention to exhaust the resources of the devices. The second category consists of privacy attacks that lead to loss of confidentiality due to unauthorized access. Finally, in [9], the authors review the use of blockchain-based solutions in the healthcare domain for mitigating security and privacy concerns. However, the focus of this paper has specifically been on only three HD stages, data collection, data sharing, and data storage.

Contrary to the existing surveys, our survey provides a holistic view of HD privacy by identifying a granular set of seven stages of the HD lifecycle and classifying various challenges and corresponding solutions for HD privacy along the identified stages. Understanding when and why these challenges happen is crucial for progressing the current state-of-the-art in ensuring HD privacy. To our knowledge, there is no previous study providing a comprehensive systematic survey of different challenges and relevant solutions related to ensuring the privacy of HD across its lifecycle. Table 1 provides a comparison of our survey with the existing survey studies. The comparison shows that, unlike existing surveys, our survey covers a complete HD lifecycle, provides a structured representation of HD privacy concerns across different lifecycle stages, and discusses open problems and future research directions in the domain of ensuring HD privacy in its whole lifecycle.

The key contributions of this paper are as follows.

- We identify a granular set of HD *lifecycle stages*, more detailed than considered previously in related work.

- We recognize a set of *privacy concerns* imposed on HD by different actors interacting with HD.

- We analyze a vast number of *techniques* proposed for improving the privacy of HD data.

- We triangulate between HD lifecycle stages, privacy risks, and techniques, as shown in Figure 1, and establish a detailed taxonomy of existing work proposed for addressing particular HD privacy concerns at a particular HD lifecycle stage.

- We provide a set of open problems and future research directions for advancing the state-of-the-art on improving HD privacy management.

The rest of the article is organized as follows. Section 2 discusses a set of privacy concerns for HD, with a special focus on different actors interacting with the data and posing risks to its privacy. Section 3 gives an overview of the existing privacy-preserving techniques. Section 4 performs a detailed review of the state-of-the-art efforts in addressing the privacy issues of HD. Furthermore, this section introduces a taxonomy that classifies the existing privacy-preserving techniques according to the relevant concern and the lifecycle stage it occurs at. Finally, Section 5 discusses the key insights of our review study and provides future research directions to improve privacy preservation for HD.

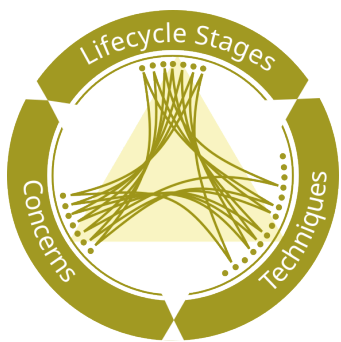

**Figure 1:** Triangle of HD Management establishes a mapping between privacy concerns, lifecycle stages they occur at, and applicable privacy-preserving techniques.

| Paper | Creation | Storage | Access | Sharing | Linking | Learning | Destruction | Concerns | Open Problems |
|---|---|---|---|---|---|---|---|---|---|
| [151] (2022) | | | | | | | | ■ | ■ |
| [40] (2022) | ✓ | ✓ | ✓ | | | ✓ | | ■ | □ |
| [86] (2021) | | | | | | | | ■ | ■ |
| [4] (2021) | | | | | | | | ■ | ■ |
| [89] (2021) | ✓ | ✓ | ✓ | | | | | □ | ■ |
| [71] (2021) | | | | | | | | ■ | □ |
| [57] (2020) | ✓ | ✓ | ✓ | ✓ | | ✓ | | □ | ■ |
| [6] (2018) | ✓ | ✓ | ✓ | ✓ | | ✓ | | □ | □ |
| [30] (2018) | ✓ | ✓ | | | | | | □ | □ |
| [10] (2017) | | ✓ | | ✓ | | ✓ | | ■ | □ |
| [21] (2016) | ✓ | | | | | | | □ | □ |
| [22] (2015) | ✓ | | | | | | | □ | □ |
| [155] (2016) | ✓ | ✓ | ✓ | | | | | □ | □ |
| [87] (2022) | ✓ | | | ✓ | | ✓ | | ■ | ■ |
| [9] (2022) | ✓ | ✓ | | ✓ | | | | ■ | ■ |
| [91] (2022) | ✓ | | | | | | | ■ | ■ |
| **This paper** | ✓ | ✓ | ✓ | ✓ | ✓ | ✓ | ✓ | ■ | ■ |

**Table 1**
Comparison with related work. The 'Concerns' column denotes the surveys that include a structural representation of privacy concerns. The 'Open Problems' column denotes the surveys that identify open problems in HD privacy management across its whole lifecycle.

## 2. Privacy Challenges of Health Data Lifecycle

In recent years, there has been a growing interest in data-driven approaches to managing HD, including Artificial Intelligence (AI) for processing HD. The benefits of such approaches are varied and many [18], however, the data-driven medical technologies may introduce higher levels of risk to HD privacy compared to traditional medical technologies [14]. This may happen because of bias inherent to AI-based decision-making [93] as well as due to the black-box nature of some AI approaches, such as deep learning [97]. Consequently, human medical professionals cannot easily supervise the use of HD to ensure there are no violations of data privacy by different actors that interact with the data throughout its lifecycle.

Next, we discuss different actors interacting with HD, posing risks to data privacy (Section 2.1). Then, we describe seven stages of the HD management lifecycle (Section 2.2). Finally, we present different HD privacy concerns (Section 2.3) and map them to the lifecycle stages in Table 2 and Table 4.

### 2.1. Actors Interacting with Health Data

There are several actors who interact with HD throughout its lifecycle, and who may consequently pose risks to HD privacy. These actors can be categorized as i) Owners, ii), Custodians, iii) Borrowers, and iv) Auditors.

HD is often owned by the patient. However, the data can be kept in the custody of HIs. We may define the term ***Owner*** as the entity that has the legal right to give consent about the application of the data, and ***Custodian*** as the entity that oversees the storage and use of data. A custodian acquires consent for storage from the owner and stores it in some centralized or distributed storage facility. The human entities in the jurisdiction of the custodian may or may not access the records depending on their level of authorization. With the consent of the owner, the custodian may choose to share the HD with some other human entities outside their jurisdiction. Due to legal restrictions, external entities are supposed to return/destroy the data after the purpose of the data application is over. Hence, the lifetime of data with the external entities is less than its lifetime with the custodian or the owner. We may generalize the external entities as ***Borrowers***, who borrow the HD owned by the owner from the custody of a custodian. The custodian may also transform the data by stripping off identifying information that is not essential for the purpose of borrowing. In that case, the borrower may independently take decisions regarding destruction and return. If the borrower borrows the HD directly from the patient then we may consider the patient to be the owner and custodian simultaneously. Finally, ***Auditor*** audits the use of data by different actors to ensure no violations of data privacy have happened. An auditor can be internal or external to the custodian organization. An internal audit mainly targets the organization's management while an external audit targets third parties. The actors are illustrated in Figure 2.

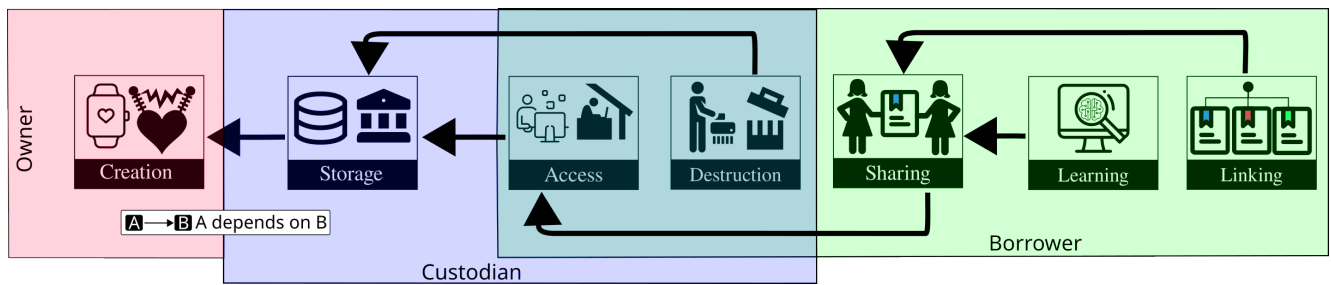

**Figure 2:** HD lifecycle stages and Actors interacting with HD.

## 2.2. Health Data Lifecycle Stages

We identify seven stages of a typical HD lifecycle, illustrated in Figure 2. Through these stages, different actors interact with the HD for different purposes. The actors perform a process to transform HD from one stage to the other. These processes require the HD to be in a particular stage. Hence, we characterize stages by their dependency on HD being in another stage. The arrows in Figure 2 denote the "depends on" relation. For example, for an HD to be in the Storage stage, it has to be collected first, which happens in the Creation stage. Consequently, we say that the Storage stage depends on the Creation stage. Additionally, some stages, e.g. Access, Destruction may happen while HD is with the Custodian as well as with the Borrower, contributing to the overlap in Figure 2. We now present an overview of each of these stages in the context of privacy and security.

**Creation:** In the Creation stage, HD is collected from the human entities through one or more intermediate devices, and health information is created. This process may involve implanted devices, wearable devices, IoT, smart home devices, etc. HD transmissions in this stage may be vulnerable to eavesdropping and manipulation. Therefore, in this stage, it is essential to ensure secure transmission of HD. A timeline of malfunctions and S&P threats of medical devices has been presented in [30]. The study spans from malfunctioning of healthcare instruments to security vulnerabilities caused by external intruders. The authors mention that since 2006 it has become a mainstream concern to secure the privacy of HD against unintended intruders. Efficient cryptographical techniques can be employed to protect the data while in transit. However, battery-powered wearable and IoT devices may not function efficiently if they need to perform expensive computations frequently.

**Storage:** After creation, HD is stored temporarily or permanently. Data may be stored in the custody of a HI, or in the custody of a patient, or even in intermediate devices connected to the internet. If the storage is compromised, then it may lead to data confidentiality issues which compromise the privacy of the human entities owning that stored data. While encrypted storage might lead to a solution, it may also degrade the accessibility of the data, e.g. searching, indexing, etc.

**Access:** Once HD is stored, access to it is often regulated by the custodian through some access policy. Failure in securing access to that storage may contribute to catastrophic data confidentiality issues. A 2016 attack on World Anti-Doping Agency [108] released confidential information about an exception on the consumption of banned substances ("Therapeutic Use Exemptions"), affecting 29 athletes including 10 American and 5 British, and many other athletes from Denmark, Russia, Poland, the Czech Republic, and Romania.

**Sharing:** HD sharing is often essential for multiple purposes, such as accessing the medical history of a patient, which may be in the custody of a different

HI, thus improving the efficiency of diagnosis, treatment, and follow-up. Furthermore, HD can be made available for research and other secondary purposes to third parties. Several approaches are taken to ensure the privacy of the patients while sharing HD. Encryption-based techniques are often used to ensure that the shared document can only be viewed by the intended actor. Some external researchers may get HD stripped out of the identifying information or even an anonymized dataset. Several vertical partitioning and anonymization schemes are proposed in multiple research papers.

- **Linking:** Sometimes it is necessary to combine data from different sources related to the same patient to create new data, thus reducing the cost of new data collection. For example, multiple medical records of the same patient scattered in multiple HIs over a long period of time can be linked together for a more complete diagnosis purpose. Anonymization techniques are often used in combination with hashing to ensure the privacy of the patients involved in HD.

- **Learning:** To infer new knowledge in the healthcare domain, it may be essential to gather and analyze a large amount of HD often using machine learning techniques. The process of learning involves handling large datasets obtained from multiple organizations. A malicious human entity involved in that process may be curious about healthcare information about a patient. HIs might perform a sufficient transformation of the data before it can be safely used for learning. At the same time, machine learning algorithms should also be capable of handling such processed data.

- **Destruction:** Once the purpose of HD is finished, it has to be destroyed to conform to the regulations and policies. If the data are left without a purpose and forgotten, then the security mechanisms that were applied to it might not still be in place, because those security mechanisms were developed for a particular purpose. This makes that data exposed to attacks because the intruders might remember the existence of that data. Therefore, it is essential to secure HD and ascertain the privacy of the patients involved in the process.

Across these stages, different actors deal with HD for different purposes. However, the past events of adversarial attacks raise several concerns that have to be dealt with in order to ensure the security of the HD and the privacy of the patients. In Creation stage, it has to be ensured that HD is not being collected by a malicious user or that the device malfunctioning does not lead to security or privacy threat to the patient. We have also seen events of data leaks while HD is being stored and accessed by human actors inside the jurisdiction of the custodian. Sharing and linking HD may also lead to HD being accessed by unintended users. Even if HD is not fully compromised, some information can still be gained by the adversaries by correlating other information available from multiple sources. In the linking as well as the learning stage, HD is vulnerable to such threats. Furthermore, the solutions (e.g. cryptographic, hashing-based, etc.) that are often adopted for such issues may obstruct the usability and accessibility of HD. Such solutions create problems in the learning stage which raises new concerns about the accessibility of HD. Finally, once HD finishes all or some of these stages and its purpose is over, it has to be destroyed so that no unintended users can make use of it.

### 2.3. Privacy concerns for Health Data

Depending on the characteristics of HD, there are different types of threats and concerns for HD privacy. HD going through all the stages of its lifecycle may encounter various threats at various stages. Some of those threats are generic to any type of data, such as Network Attacks and Storage Attacks. Some attacks, which are also applicable to other types of data are of special interest in healthcare, such as Reconstruction Attacks and History Manipulation Attacks. In [155] a variety of security attacks are discussed in the context of healthcare. However many of the attacks described in that paper may relate to the same security concerns. For example, the three categories explained in the paper Masquerade attacks, Body-coupled communications, and Attacks on wearable and implantable medical devices are all related to impersonation. Similarly, the primary motive of the data injection attack is to cause device failure, thus Denial of Service. In this paper, we categorize the threats on HD as mentioned in the literature into concerns. Next, we discuss each of these categories of attacks.

#### 2.3.1. Networking Concerns

In several lifecycle stages, HD is in transit from one device to another, or from one institution to another. In such circumstances, many types of network attacks can endanger the privacy of the patient. There have been several past incidents related to such attacks. For example, implantable Medical Devices (IMDs), Implantable Cardioverter Defibrillators (ICDs), pacemakers, Neurostimulators, and Implantable Drug Pumps are automated battery-powered computing and communication systems that monitor the patient's body vitals and provide automatic therapies. HD is generated by the implanted or wearable smart health devices or by the monitoring systems of HIs. The generated data is then transmitted to a receiver that either analyses the data or forwards that data to a storage system. When the data is transmitted wirelessly it is prone to network attacks. We can broadly categorize the Network attacks into the following four categories.

- **Impersonation:** The adversary pretends to be a legitimate user and tries to send false HD or commands. Both implanted and wearable medical devices are prone to this type of attack. Attackers use foreign transmitters to send malicious packets to the nodes inside the body area network. Such a false HD not only endangers patients' privacy but also is a threat to patients' security.

Also, a foreign device may send malicious signals to the other nodes to collect data from them.

**Eavesdropping:** A passive attacker can eavesdrop on the RF communication and an active attacker may impersonate and control medical devices to alter the intended therapy. In [55] the authors analyze the RF transmissions by the ICD and commercial ICD programmers and were able to decode the signals. Hence a passive adversary can eavesdrop on the communication using commercially available equipment. An active adversary may generate arbitrary RF traffic and spoof a commercial programmer. The monitoring system may be overwhelmed by a large number of spam messages and the legitimate messages may become lost [31], which can have fatal consequences. In [75] the authors demonstrate the possibility of active and passive attacks on popular glucose monitoring and insulin delivery systems available on the market.

**SPoF:** Modern HD management involves the cooperation of multiple devices across the network. A failure in a device on which multiple entities are dependent may result in the inaccessibility of HD. Moreover, it may lead to an advantageous situation for an adversary who intends to impersonate a legitimate user or eavesdrop on the network communication.

**DoS:** Denial of Service (DoS) refers to an attack scenario when an adversary tries to disrupt the system by interacting with the system too frequently or by sending false data[155] and keeping the system busy. The resource-constrained battery-powered implantable devices can be victims of such attacks. The attacker may keep the device busy in communication while draining its battery. Moreover, an external intruder may disrupt the communication between an implantable device and a valid caregiver [104].

### 2.3.2. Storing Concerns

Once the data is transmitted, it is stored in the custody of some HI. That institution has the responsibility to secure HD from unintended or unethical access or data leaks. Indian legal systems require every physician to maintain the medical records pertaining to his/her indoor patients for a period of 3 years and "If any request is made for medical records either by the patients/authorized attendant or legal authorities involved, the same may be duly acknowledged and documents shall be issued within the period of 72 hours" [2] [1]. The HIPAA regulations require up to six years of record retention period [88]. The GDPR and UK Data Protection Act mandates medical record retention for a minimum of eight years. There are legal restrictions on the usage of data put by the HIPAA and GDPR. However, in 2015, Radio Shack wanted to sell user data to pay off its debt while Apple and AT&T filed a motion to prevent [61][60]. Various loopholes in the legal restrictions of HIPAA are identified in [12]. Snowden's revelations regarding U.S. intelligence agencies' surveillance of commercial Internet services endanger the data sovereignty of the parties involved in the transaction of HD[12].

Hence, it is shown that HD is a valuable commercial asset that many extra-institutional or even intra-institutional entities would like to have access to even when it is not intended. HD in storage is vulnerable to several attacks. If the HD storage is not protected then it can be accessed by internal or external adversaries. There have been several techniques used in the literature to ensure that HD is being accessed by the intended users only.

**Confidentiality:** From 2010 till 2019, there have been 3253 instances of health data breaches [114] related to HD stored in paper or electronic devices, e.g. laptops, desktops, email Electronic Medical Record (EMR), etc. It has also been found that some health applications have transmitted information to third parties who are often data brokers, who aggregate the data and sell it in the market. Moreover, it has also been reported that hospitals, insurers, and grocery retailers have collaborated to derive health risk profiles from buying patterns of individuals [102]. On the one hand, state security and "terror" is used as an excuse to justify surveillance on individuals [120]. On the other hand, the accumulated data is available for sale. It has been found that the estimated price of a patient's health record is $50, whereas social security numbers and credit card information are sold for $3 and $1.50 only [8].

**Manipulation:** An National Health Services (NHS) Trust deliberately deleted 90,000 emails that were linked with two deaths and under-staffing that were critical to a legal case brought by a whistleblower [52]. In a 1996 Indian court case, the court charged the HI for negligence and manipulation of HD [128]. A patient in Kentucky, USA was not diagnosed as a Cancer patient, as she received a false test report, however, it was later found out that she had stage 4 cancer [117]. After a lawsuit Forensic Experts found that electronic records were edited to delete the erroneous letter claiming she was free of cancer [43] [56]. In Essex, England police investigation revealed that records of 22 out of a sample of 61 patients were manipulated to conceal the fact that they have faced extensive delays in their cancer treatments [39].

**Accessibility:** Once the data in storage is transformed, due to confidentiality, manipulation, or some other concerns explained later, the usability of the data may be reduced. It requires special techniques in order to make that transformed data usable. For example, a dataset may be encrypted due to confidentiality concerns. However, it may be necessary to keep it accessible for search operations. It is often necessary to keep the encrypted data accessible in the absence of the patient, in emergency situations. Additionally, different research institutions may use the data to infer new information while keeping the data encrypted. Such accessibility concerns are often planned in advance and taken care of while storing the data.

**Traceability:** Human entities often have access to HD at different data lifecycle stages. These human entities may gain information and take further actions on HD, in spite of not being the owner of the data. Such actions often require consent from the actual owner of the HD. Additionally, these entities have to be held responsible for their actions and suggestions. Hence it is essential to keep track of the history of the events that are associated with the HD.

*2.3.3. Reconstruction Attacks*

To protect the privacy of the patients, HD is often transformed in a way that disassociates the patient's identity from the HD. In [126] it was found that about half of the US population can be uniquely identified from a subset of quasi-identifiable information using 1990 census data. Similar studies on 2000 census data have found similar risks [50]. Under HIPAA regulations, HD is deemed to be de-identified if it does not contain any of the 18 specified data elements [41]. De-identification reduces the risk of data breaches as the adversaries do not get the ID information associated with the SI information. However, it is still vulnerable to Reconstruction attacks where the statistical information from multiple medical documents is used to re-identify the patient. Canadian Federal Court has decided that the province of residence information has to be excluded from HD in the adverse drug event dataset to protect individuals from being re-identified using statistical information [41]. In [80] the authors have investigated de-identified New Zealand HD retrieved from three Primary Health Organizations and found a significant risk of re-identification. There have been 55 attempts of successful re-identification [58]. Often multiple institutions share their HD to find out the intersections, using which they can build a more complete patient profile. However, such sharing of information is also prone to similar vulnerabilities from external or internal adversaries. To address similar issues with Court Decisions the authors conduct a study on anonymization of German Court Decisions[33]. Experiments have been carried out to de-anonymize court decisions anonymized by law students. The study found that 38% of the anonymized strings can be de-identified.

**Disclosure:** Unlike Confidentiality when unintended access is performed by an adversary, there can be instances of unintended exposure while sharing HD intentionally. For example, the dataset has been shared with some external researchers for knowledge discovery. The data owners (usually the patients) have given consent for that specific type of research. However, that data can also be used for some other purposes for which the patients have not given consent. The external researchers may also gain personal information about the patients from that HD. Such application of the data will be unethical and is not permitted by law in various countries. Many authors propose techniques such that the data does not disclose any other properties other than the properties asked for.

**Re-Identification:** During data disclosure, the shared data is used for some unintended application. However, there is another risk of privacy infringement, even when the HD is not shared, even if HD does not include the identifying information of the patients. The anonymized or pseudoanonymized medical information in the HD can be co-related with some publicly available information and such co-relation can identify some individuals. For example, a dataset containing only a record referring to one diabetic patient and including skin color, age, and location information, but including no personal details like phone number, name, etc. can still be used to identify that person.

Table 2 maps different HD lifecycle stages to relevant concerns and Table 4 gives a systematic overview of the state-of-the-art techniques addressing different HD privacy concerns involving different actors interacting with HD at different stages of HD lifecycle.

## 3. Overview of Privacy Preserving Techniques

In this section, we provide a brief overview and describe a general working of all the techniques that we have encountered in our survey. A review of how are these techniques applied to preserve the privacy of HD is given in Section 4. Although many of these techniques are used for protecting the privacy of the parties involved in the HD lifecycle, many of these techniques are also used for many other generic purposes, such as secure communication, or knowledge discovery. However, these techniques are often applied to problems associated with the privacy of HD in the literature.

### 3.1. Cryptographical Techniques

The process of transforming the plaintext into ciphertext is called encryption, and reversing it using a trapdoor is called decryption. There are several types of encryption techniques such as Symmetric encryption, Asymmetric encryption, Cryptographic hashing, etc. *Symmetric Encryption* is an encryption technique in which the same secret is used for encryption and decryption. Symmetric encryption algorithms include *AES*, *DES*, *Blowfish*, *RC4*, *RC5*, and *RC6*. On the other hand, *Asymmetric Encryption* uses a public key to encrypt and a private key to decrypt. There are various algorithms to perform Asymmetric encryption, such as *RSA*, *Elliptic Curve Cryptography*, *ElGamal*, etc. Both of these encryption techniques are used for communication and document exchange in multiple domains including healthcare. Although, symmetric encryption is more efficient for exchanging a large amount of data [159]. It requires the communicating parties to be aware of the secret key, which has to

| Stage | Concern | References | Stage | Concern | References |
|---|---|---|---|---|---|
| Creation | Impersonation | [147] [103] [110] | Sharing | Impersonation | [130] |
| | Eavesdropping | [72] [144] [118] [76] | | Confidentiality | [148] [129] [130] |
| | Confidentiality | [38] [26] | | | [26] [99] [78] [59] |
| | Manipulation | [38] | | Manipulation | [78] [59] |
| | Traceability | [38] | | Traceability | [59] |
| Storage | SPoF | [78] [145] [67] | | Accessibility | [156] |
| | Confidentiality | [148] [100] [38] [129] [130] | | Re-Identification | [26] |
| | | [26] [78] [90] [149] [35] | Linking | Confidentiality | [35] |
| | | [59] [42] [70] [145] [67] | | Re-Identification | [84] |
| | Manipulation | [152] [156] [90] [35] [78] | | Disclosure | [84] [19] [24] [113] |
| | | [38] [59] [145] [67] | | | [28] [131] [112] |
| | Traceability | [152] [156] [90] [38] [59] [145] [67] | Learning | Confidentiality | [47] [59] |
| | Accessibility | [152] [145] [67] | | Manipulation | [156] [35] [59] [119] |
| | Re-Identification | [76] [143] [148] | | Traceability | [152] [156] [47] [59] |
| | Disclosure | [143] | | Accessibility | [152] [115] [148] |
| Access | Confidentiality | [148] [100] [129] [26] [99] | | | [54] [49] [23] |
| | | [78] [90] [42] [70] | | | [47] [156] [94] |
| | Manipulation | [156] [90] [78] | | Disclosure | [54] [94] [49] |
| | Traceability | [156] [90] | Destruction | Manipulation | [152] |
| | Accessibility | [90] | | Traceability | [152] [46] |

**Table 2**
State-of-the-art techniques addressing HD privacy concerns occuring at different HD lifecycle stages be communicated securely. There are many *Key Agreement* protocols, such as *Diffie Hellman* that securely compute the same secret key on both ends by exchanging randomly generated secrets using Asymmetric encryption. However, in the context of healthcare devices, alternative techniques are often applied to minimize the computational overhead. Such techniques use commonly observed phenomena, e.g. ECG or RSS to determine a secret to be used for further communication.

Cryptographic signatures provide security against tampering with the original document and prove authenticity. However, patients interested in sharing partial HD also need to fulfill the same security requirements for partial documents. Signature schemes like *Content Extraction Signature (CES)* [121] are used for extraction of a verifiable partial document without requiring the re-signing of the extracted parts, which is often used in the literature. Also, a patient may prefer to remain anonymous while sending HD, while the receiving entity has to verify the integrity of the transmitted data. Techniques like Ring Signature [81] are often used for these types of problems.

Secret Sharing techniques are used to permit decryption by collaboration of multiple entities. Techniques such as *Samir's Secret Sharing* or *Blakley's Secret Sharing* are often used in such scenarios, to make HD accessible by multiple actors in the Healthcare system. Security frameworks like *Identity-based Encryption (IBE)*, *Attribute-based Encryption (ABE)*, *Key Policy-based ABE (KP-ABE)*, or *Ciphertext Policy-based ABE (CP-ABE)* use these techniques to provide a system of access control, which is often used for ensuring S&P in the literature.

Although encryption makes HD secure from unintended access, it does not allow any qualitative or quantitative analysis on encrypted HD. Hence, various *Homomorphic Encryption* techniques like *ElGamal*, *Paillier* are used for encrypting HD in such a way that permits algebraic operations. However, most of the widely used Homomorphic Encryption algorithms allow only a subset of operations. Therefore, different *Fully Homomorphic* schemes are also applied on HD, but that increases computational overhead [136] and often makes it difficult to implement.

### 3.2. Hashing Techniques

Hashing is a technique of mapping the input items into a finite set which is often used in efficient storage, retrieval, and computation. Certain types of hashing techniques are often used in network communications, for the integrity and authenticity of a message. In the context of S&P of HD, hashing is also used for similar purposes. *Cryptographic hashes* like *MD5*, *SHA1*, *SHA256*, and *SHA512* are used for calculating fixed-sized irreversible digest to ensure the integrity and to defend against the manipulation of HD. However, a small change in the input contributes to a huge change in the digest, which makes it difficult to co-relate two inputs using the two digests, which is often required in HD. Hence, *Locality Sensitive Hashing* techniques are used in literature to compute digests in such a way that preserves the distances between the inputs, which is the HD originating from health sensors. *Bloom Filter*s on the other hand use combination of multiple hashing algorithms and provide a space-efficient data structure that encodes multiple elements of a set into a single-bit string. Such techniques are widely used for finding an approximate similarity between HD, without disclosing private information. Different variants of Bloom Filters are used for encoding different hierarchical data (e.g. decrease codes) in HD.

### 3.3. Secure Computation

HD is generally stored in the custody of an Institution. Multiple such custodians may want to perform a collaborative analysis. However, exchanging HD outside their premises may violate privacy concerns. Therefore, the computations have to be performed in-house. However, the correctness of the computation has to be ensured for the other custodians to use the intermediate results. That leads to various applications of *Verifiable Computing* techniques like *zk-SNARK* [111], *Garbled Circuit*[150], etc. *Multi-party Private Set Intersection (MPSI)* [73] techniques are also used for finding the intersection of privately held information. In order to ensure privacy, hashing, and encryption techniques are often incorporated with MPSI in literature.

### 3.4. Ledger Techniques

Ledger techniques are generally applied to provide security against manipulation of HD and relevant information regarding operations on HD. Ledger techniques also provide protection against a single point of failure and manipulation through the collaboration of multiple parties. Such techniques primarily use *Markle Tree* in which every non-leaf node is labeled with the cryptographic hashes of its child nodes, whereas the leaf nodes are labeled with the cryptographic hash of their contents. So, manipulating one node in the tree invalidates the entire tree. Similar to the Markle Tree, a *blockchain* consists of blocks, each containing one or more transactions. A block in the blockchain is linked with its previous block using cryptographic hashes. The incorporation of a consortium of multiple healthcare entities in the blockchain increases trust and reduces the threat of a single point of failure. *Smart Contracts*, on the other hand, are small programs in the blockchain that are executed by the distributed system. The consortium verifies the input and the output of the Smart Contract. Such systems are often used in the literature to protect HD from being manipulated and to maintain the history of operations performed on HD. However, it is difficult to validate the correctness of such blockchain-based systems [82].

### 3.5. De-Identification Techniques

HD has a purpose depending on the lifecycle stage it is currently in. These purposes often do not require the identity of the patient with whom the sensitive medical information is associated. De-Identification techniques are used to transform a data or dataset into another dataset that serves the purpose while removing or obfuscating the identifying parts. There are several techniques that are used for this purpose. Information in HD can be categorized into three categories: Identifier (ID), Quasi Identifier (QI), and Sensitive (SI). Identifiers include information that can be used to directly identify a person, e.g. name, social security number, address, etc. Quasi Identifiers are generic descriptions of a person that cannot be used directly to identify a person. However multiple such Quasi Identifiers can be used with other publicly available information to identify a person e.g. color of skin, height, weight, etc. Sensitive Information is the medical information associated with a patient that has to be protected from being identified. In [109], the authors propose a machine learning-based approach to re-identify persons from HD, which is generated from the wrist-worn motion sensor data. Re-Identification attacks are not only limited to tabular data. In [92], the authors propose deep learning-based techniques to re-identify patients based on their chest X-rays. The proposed system is able to identify whether two X-ray images belong to the same patient or not, with high accuracy. ECG-based patient identification has been performed in [48] using CNN. The model attained an accuracy rate of 94.56%.

Different techniques for the anonymization of HD have been used in the literature. *Pseudoanonymity* can be achieved by replacing the personal identification of patients with some pseudonyms that do not directly identify the patient involved. However, the HD can still be statistically co-related, which may lead to the gain of information about the patient's identity. *Vertical partitioning* techniques are used to split the schema and isolate a subset of information from HD and store it separately. *K-Anonymity*[127] based anonymization techniques are often used in the literature to ensure that, given any record, there are at least $k-1$ other records in the dataset that are indistinguishable from the given record. In order to make such a transformation, the QI values are often swapped, randomized, generalized, or suppressed. Generalization of QI value implies replacing the original value with some superset of information, e.g. replacing age 25 with a range 20-30. Whereas suppression implies completely removing the original value, e.g. replacing it with '*'. However, this does not provide a strong guarantee of privacy preservation. Because if the SI information associated with the group of k records is less diverse, it may make it easier to identify the person with some probability. For example, if a group of k records is associated with the same disease, then it is obvious that if a victim patient falls in that group he or she must be associated with that disease. In [79] the authors introduce two Privacy principles. Positive disclosure is when an adversary can identify the value of a sensitive attribute from a k-anonymized dataset. Negative disclosure is when the adversary can eliminate some possible values of a sensitive attribute. The paper also proposes *l-diversity*[79] that diversifies the sensitive information in every group in a k-anonymized dataset. However, it is often very difficult[77] and sometimes impossible to achieve *l-diversity*. For example, if the sensitive attribute does not have enough diversity in the whole database, it is too difficult to diversify it in every group. Hence, *t-closeness* is proposed as another metric of privacy. It is based on the distance between the distribution of the attribute in the group and its distribution in a group.

### 3.6. Learning Techniques

Big datasets are often used for analysis so that new knowledge can be learned and applied to administrative, political, or marketing decision-making systems. In the literature of S&P of HD, predictive models are often built with techniques like *Auto regression*, *Decision Tree*, or neural network-based techniques e.g. *Artificial Neural Network*,

*Long Short Term Memory (LSTM)* etc.. *Clustering* techniques like *k-means* have been used to partition a dataset into a finite number of subsets, each containing similar HD. However, in the case of HD, it is necessary to preserve the privacy of the patients while using learning algorithms. Hence, many works incorporate *Homomorphic encryption* into learning techniques. Moreover, if the HD is already encrypted, the searching techniques become difficult to apply on ciphertext. Therefore, search techniques like *Public Encryption with Keyword Search (PEKS)* are used for efficient search over HD.

### 3.7. Content Processing Techniques

A recent study has found vulnerabilities in the DICOM[1] protocol [138], which puts the confidentiality of a large number of users at risk. In [83], the authors perform a MiTM attack on the CT scanner machine to get the imaging contents generated from the machine. The paper proposes a machine learning-based technique for injecting or removing features from the 3D imaging HD. Such techniques can be used by attackers to manipulate data-at-rest as well as data-on-move. For imaging HD, Steganography techniques such as *Reversible Data Hiding (RDH)* are used to hide private data inside medical imaging. For textual contents, *q-gram* based techniques are often combined with *Bloom Filter* to enable privacy-preserving fuzzy comparison over HD.

## 4. Privacy of Health Data Lifecycle: Taxonomy and Review

In this section, we review the state-of-the-art approaches for ensuring the S&P of HD. The review is structured per privacy concern type. Furthermore, we propose a taxonomy of major techniques relative to the privacy concerns they address. The taxonomy is presented in Table 3, with techniques shown on the left and concerns on the top. In Table 4, we associate the technical approaches with the concerns and the techniques used to solve them. We also mention the stages that the works span along with their brief description.

In our review, we have found the heavy application of cryptographic techniques in different HD lifecycle stages addressing one or more S&P concerns. For example, we can see from the taxonomy that nearly all concerns are addressed by some cryptographic technique. SPoF and Traceability are often addressed by blockchain, which is not a cryptographic technique on its own, however, it uses cryptographic techniques to ensure the immutability of the ledger. The first two network concerns, Impersonation and Eavesdropping are primarily encountered in the Collection stage of the HD lifecycle, as shown in Table 2. However, blockchain is often used as a failsafe storage system that can tolerate SPoF while making the storage traceable at the same time. A comprehensive overview of the state-of-the-art techniques addressing different HD privacy concerns involving different actors interacting with HD at different stages of the HD lifecycle is shown in Table 4.

### 4.1. Literature Search Protocol

We followed a four-step literature search protocol while searching for papers that propose technical solutions against the S&P concerns of HD. First, we focused on the articles whose abstracts include the following keywords: "health data" or "healthcare data", privacy", "security", "privacy-preserving", "vulnerability", "identification", "de-identification", "health data life cycle", "anonymization". We restricted our search to conference and journal articles written in English and published between 2015 and 2023 in the field of computer science. We used the platform Scopus as a source of high-quality research literature on our topic of interest. The search returned 125 articles. In the second step, two reviewers performed preliminary paper selection by analyzing the articles returned the by search in Step 1 using Exclusion and Inclusion criteria. Specifically, we applied the Exclusion and Inclusion criteria to the paper title, keywords, and abstract. The Inclusion criteria were methods, algorithms, and case studies for health data privacy preservation. The Exclusion criteria were surveys and mapping studies, position papers, short abstracts, editorials, and panels. The Inclusion and Exclusion criteria reduced the number of articles to 88. In the third step, we performed a detailed assessment of the papers selected after the Inclusion and Exclusion criteria were applied. Two reviewers read the papers to select those that fully cover the objective of the study and that are of high quality and present mature ideas. This step returned 45 papers. Finally, we applied the snowballing technique to identify additional important articles relevant to our survey. Specifically, we used a reference list of a particular paper to find additional interesting papers, without limiting the time when the paper was published. After snowballing, we had a final list of 52 papers that were included in the study. In the following subsections, we discuss the papers that propose technical solutions the addressing S&P concerns of HD.

### 4.2. Network Concerns

Impersonation and Eavesdropping are the two important concerns in the domain of implanted and wearable devices. The major problems identified by the technical works in this field are key establishment and protection against adversaries in the network. In [103] a proximity-based cryptographic protocol is proposed for securing communication between implanted devices and the reader. The reader works as a prover that must prove its proximity which is verified by the implanted device The authors evaluate the effectiveness of the proposed work by experimenting with a transmitter implanted inside the beef. The effectiveness of the proposed protocol depends on the accuracy of distance measurement. The experiments conclude the error is ±1.78cm through the air and ±0.01cm through 2cm meat and 1cm air. In [118] [147] the authors use RSS fluctuations to differentiate the on-body and off-body sensors to protect from impersonation attacks. In [147] Auto regression and Long Term Short Term Memory (LSTM) based techniques are used for signal

---

[1]Digital Imaging and Communications in Medicine (DICOM) is the international standard for medical images and related information (ISO 12052)

| Techniques | Impersonation | Eavesdropping | SPoF | DoS | Confidentiality | Manipulation | Traceability | Accessibility | Re-Identification | Disclosure |
|---|---|---|---|---|---|---|---|---|---|---|
| **Cryptographic** | | | | | | | | | | |
|   **Access Scheme** | | | | | | | | | | |
|     Secret Sharing | | | | | | | | | | |
|     **Frameworks** | | | | | | | | | | |
|       CP-ABE | | | | | [129][78] | | | | | |
|       KP-ABE | | | | | [46] | | | | | |
|       IBE | | | | | [124] | | | | | |
|       Public Key Cryptography | | | | | [38] | | | | | |
|   **Signature** | | | | | | | | | | |
|     Content Extractable | | | | | [78] | [78] | | | | |
|     **Source Obfuscating** | | | | | | | | | | |
|       Ring Signature | | | | | | [38] | | | | |
|       Blind Signature | | | | | | [35] | | | | |
|     Chaoatic | | [15] | | | | [15] | | | | |
|   **Key Agreement** | | | | | | | | | | |
|     Secret Exchange (DH) | [103] | | | | [130][38] | | | | | |
|     **Phenomenon based** | | | | | | | | | | |
|       RSS based | [118] | | | | | | | | | |
|       ECG based | | [144] [11] | | | | | | | | |
|   **Homomorphic** | | | | | | | | | | |
|     Pallier | | | | | | | | [47] [54] | | |
|     Additive Elgamal | | | | | | | | [36] | [84] [26] | [36] [84] |
|     Fully Homomorphic | | | | | [125] | | | [125] | | |
|   **Symmetric** | | | | | | | | | | |
|     AES | | [72] | | | [148] [100] [124] | | | | | |
|     ARX | | | | | [38] | | | | | |
|     RC5 | | [55] | | | | | | | | |
|   **Asymmetric** | | | | | | | | | | |
|     RSA | | | | | [35] [70] [148] | | | | | |
|     Elliptic Curve | | | | | [90] | | | | | |
|   Post Quantum (NTRU) | | | | | [26] | | | | | |
| **Hashing** | | | | | | | | | | |
|   **Bloom Filter** | | | | | | | | | | |
|     Lattice Based | | | | | | | | | | [131] |
|     Double Hashing | | | | | | | | | | [112] [103] |
|      | | | | | | | | | | [84] [28] |
|     Hierarchy Preserving | | | | | | | | | | [113] |
|     CLK | | | | | | | | | | [19] |
|   Universal Hashing (SFIF) | | | | | | | | | | |
|   **Cryptographic** | | | | | | | | | | |
|     Location Sensitive | | [110] | | | | [110] | | | | |
|     HMAC | | [110] | | | | | | | | |
|     PUF | [137] | [137] | | [137] | | | | | | |
| **Secure Computation** | | | | | | | | | | |
|   Verifiable (zk-SNARK) | | | | | | [59] | [59] | | | |
|   **Multi Party** | | | | | | | | | | |
|     Garbled Circuit | | | | | | | | | | [24] |
|     Set Intersection | | | | | | | | | | [84] |
| **Immutability based** | | | | | | | | | | |
|   Markle Hash Tree | | | | | | | [46] | | | |
|   IPFS | | | [145] [67] | | [145] [67] | [145] [67] | | | | |
|   Blockchain | | | [78] | | [152] [156] [78] [90] [145] [67] | [152] [156] [90] [145] [67] | [152] [156] [90] [47] [145] [67] | | | |
|   Smart Contract | | | | | [38] [47] [59] | | [38] [47] [59] | [47] | | |

strength prediction, which is used for detecting malicious behavior. However, a completely different approach is taken in [158], which secures the network of medical devices by observing their communication from outside. The authors in [158] propose a medical security monitoring system that snoops on all the radio-frequency wireless communication to identify adversaries. Spoofing techniques can also be used to impersonate the server and make the devices transmit their data to the adversary. In [154] a Man in the Middle attack scenario has been considered where a malicious client node in the Bluetooth Low Energy (BLE) network copies all properties of the server and broadcasts fake advertising packets. The medical devices become victims and transmit the body vitals to the adversary node instead of transmitting them to the real server. The authors propose a monitoring-based solution to address and mitigate this attack. The proposed work creates a constellation of decision regions based on response times which it uses to characterize the normal behavior of the device. Metrics like false positive and false negative has been use as the metric of effectiveness in [118][147][158][154]. In [72] real-time symmetric encryption techniques have been proposed to secure the communication between medical devices. In [144] an external guardian device is used to proxy the implanted medical device. The authors use ECG-based key agreement to derive a shared secret. Some recent works [11] detects QRS complex and combines that with LFSR to generate initial keys for key agreement. With experiments the authors conclude that jamming the implanted device's transmission is more effective than jamming the adversary's transmission. They also conduct experiments to conclude the distance between the IMD and the Guardian node is close enough (2 feet) to be effective. In [15] a secure resource-efficient communication scheme is proposed for implantable medical devices that use Henon scheme-based chaotic systems to protect the messages from eavesdropping. The paper measures the robustness of the signature against statistical attacks. In [110] a framework has been proposed for preventing Man in the Middle attacks in the network of medical devices. Instead of transmitting the data, a locality sensitivity hash of the data is transmitted along with its Hash Message Authentication Code (HMAC) which makes it impossible for eavesdroppers to understand the meaning of the message or to modify it. The authors evaluate their proposal as a proof of concept system using Raspberry Pi, with the e-Health sensors platform. PhysioNet dataset is used instead of real-time data for the experiments. They use the True positive and False alarm rates as metrics for performance analysis. However application of cryptographic solutions on resource-constrained sensors may lead to increasing power consumption and battery draining. Some recent works[137] address these challenges on BAN (Body Area Network) sensors by using *Physically Unclonable Functions (PUF)* instead of using Cryptographic solutions. This relies on the fact that despite being built with the same design each hardware device may demonstrate some unique features, due to the differences in manufacturing. The proposed approach utilizes that and uses PUFs as fingerprints of the device that can be computed based on a given input. The input-output key pairs are used for authenticating the devices. Moreover, external adversaries can launch a Denial of Service (DoS) attack on an implantable device by draining its battery while keeping it busy in communication [55]. Some implanted devices also maintain a log of the transactions with external devices. The adversary may also launch a DoS attack intended to overflow the device's onboard memory [20]. Zero Power Defence (ZPD) techniques have been proposed in [55] for defending against such attacks. In [55] the authors propose harvesting radio frequency energy to perform Zero power notification and authentication to save the main battery from depletion. The notification system can wirelessly activate a piezo-element that can audibly warn the patient against security-sensitive events. The proposed authentication technique uses RC5 encryption to verify that it is communicating with an authorized programmer device. A similar approach is used in [123]. The authors in [123] propose a System on Chip (SoC) architecture that partitions the system into two modules. The module for the main implant functionality is connected to the battery. However, the security Core which is a separate module can also be powered by the energy harvested by the RF antenna. Any new communication wakes up the Security core which then authenticates and verifies the communicating external programmer. The paper proposes a security protocol for authentication that conforms to ISO/IEC 9798.

| Techniques | Impersonation | Eavesdropping | SPoF | DoS | Confidentiality | Manipulation | Traceability | Accessibility | Re-Identification | Disclosure |
|---|---|---|---|---|---|---|---|---|---|---|
| **Learnng** | | | | | | | | | | |
|   **Clustering** | | | | | | | | | | |
|     KNN | | | | | | | | [94] | | [28] |
|     K Means | | | | | | | | [54] [49] [23] | | |
|   **Regression** | | | | | | | | | | |
|     Autoregression | | | | | | | | [147] | | |
|     Decision Tree | | | | | | | | [115] | | |
|   Rule Mining | | | | | | | | [35] | | |
|   Neural Network | [147] | | | | | [119] | | [147] [46] | | |
|   **Cryptographic Search** | | | | | | | | | | |
|     PEKS | | | | | | | | [156] | | |
|     Hybrid | | | | | | | | [148] | | |
| **De-Identification** | | | | | | | | | | |
|   Verticle Partitioning | | | | | | | | | | |
|   **Anonymization** | | | | | | | | | | |
|     k-anonymity | | [76] | | | | | | | [148] [76] [143] | [143] |
|     l-diversity | | | | | | | | | | |
|     t-closeness | | | | | | | | | [148] | |
|     Distributed Randomization | | | | | | | | | [143] | [143] |
|     KNN based | | | | | | | | | [53] | |
|   Pseudonym | | | | | [99] [78] [90] | | | | | |
| **Text Processing** | | | | | | | | | | |
|   q-gram | | | | | | | | | | [112] |
| **Image Processing** | | | | | | | | | | |
|   Steganography | | | | | | [119] | | [149] | | |
| **Zero Power Defence** | | | [55] [123] | | | | | | | |

**Table 3**
Taxonomy of major techniques for preserving the privacy of HD structured according to concerns addressed.

In summary, we see a consistent pattern of using non-cryptographic alternatives to address network concerns in HD originating from medical devices. The alternatives can be categorized as an observed phenomenon, unique device characteristics, and lightweight encryption alternatives.

### 4.3. Storage Concerns

HD is often stored in the custody of some HI. That institution has to safeguard the data from being accessed by unintended users with malicious intentions inside and outside the institution. Several techniques have been used to ensure the integrity and confidentiality of HD. Cryptographic techniques are often used to protect the data even if the storage system is compromised. Cryptographic access schemes are used to ensure the data is being accessed by the intended users only. Signature schemes are used to protect the HD from manipulation. Ledger-based techniques provide immutability of HD as well as actions of HD which enforces the responsibilities of the actors involved in the HD management lifecycle. However, although these cryptographic techniques strengthen the security aspects of HD, they simultaneously make the data less accessible and usable for analysis like machine learning. Hence several cryptographic techniques are applied to the data in order to keep the data usable while ensuring the security of the data. Now we discuss the relevant researches that address the storage concerns by applying different techniques.

#### 4.3.1. Cryptography based

Cryptographic techniques are also used for storage concerns. In [99] Transparent Data Encryption (TDE) technique is used for encrypting the HD stored in the SQL Server database. With this technique, all data is encrypted before being written to the disk which is decrypted back to plaintext while retrieving. On the other hand, in [100] HD is split into multiple 256-bit fragments. Such fragments are again broken into 8 smaller chunks, out of which one random chunk is considered private and the rest public. The public chunks are protected by xor'ing with the hash of the private chunk, a secret key, and a counter. The private chunk is stored in the user's device. Even if an attacker gets access to the storage and has the secret key, and the counter, the attacker still needs to have access to the private part which is stored in the user's device. Encrypted HD may become problematic in the time of emergency. However, keeping it unencrypted may be a risk due to confidentiality concerns. The authors evaluate the execution speed for the main calculation tasks using iPhone 8 Plus. Hence in [130], a patient and doctor use Diffie–Hellman and the sibling intractable function families (SIFF) algorithm to establish a shared secret which is used for encrypting medical documents shared between them. With this proposal, a patient can form a group of other individuals who have the ability to decrypt the encrypted documents. The group may consist of pharmacists, family members, legal/insurance agents, etc. Blockchain platform Hyperledger Fabric is used for document exchanges. Authors measure throughput as a metric of the effectiveness of their proposed framework. The authors in [129] mention three levels of confidentiality of medical information that allow the owner to define confidentiality of his/her own personal health records. A secure level information can only be accessed by the emergency staff at the time of emergency. A restricted level information can be accessed by the emergency staff only if k out of predefined n trusted users grant permission. An exclusive level of information can never be accessed by the emergency staff. The encrypted medical records are stored on a server. Encryption is performed using (k,n)-threshold cryptosystem. In [70] a different approach is proposed while using the same (k,n)-threshold cryptosystem. The medical records are encrypted using the RSA algorithm. The private key, which is required to decrypt the records, is shared using (k,n)-threshold cryptosystem. Instead of giving these shares to human entities, they are stored on the server. Each of these shares corresponds to different context conditions, such as the doctor's identity, role, location, duty time, patient location, status, etc. The authors implement the proposed scheme and measure the performance as time of execution. In [42] (k,n) threshold cryptosystem is used to securely store patients' healthcare records. However, the authors identify several problems while applying the original secret sharing scheme proposed by Shamir et. al in the healthcare problem. Using the same secret for encrypting all health records is vulnerable to attacks. Once the secret key is revealed it is revealed forever. Additionally, the participants must reveal their share of the secret in order to reconstruct the secret key. Therefore the paper proposes a novel cryptographic scheme based on the original Lagrange interpolation-based threshold cryptosystem. In the proposed protocol, the participants have their secret shares with them but they do not share them. Instead, they submit a transformed value to the server. The original secret share cannot be retrieved from that transformed value. The paper also includes a variation of the proposed scheme which is used in cases of emergency, such as when the patient is unconscious and the security team can act on behalf of the patient. Cryptographic access schemes like IBE, CP-ABE are used in [153], [124] [129] [78]. In [124] the medical data is first encrypted by the sender using the symmetric encryption algorithm AES. The secret key is encrypted using IBE and shares the encrypted document along with the encrypted key. The authors implement the proposed system and use the processing time of various cryptographic operations as the metrics of effectiveness. In [153] a Hierarchical access scheme is proposed, where the Public Health Office serves as the Public Key Generator (PKG) at the highest level, and the Hospitals, and Clinics are in the lower level. The storage servers located at hospitals and clinics store the medical records of their patients only. The public storage server is responsible for storing the referral medical records. In [46] IBE is used along with a Markle Hash Tree to ensure the deletion of HD. The assured deletion algorithm proposed in [146] has been incorporated in this work. The deletion scheme updates the root of the tree which can be verified by the patient who requested for deletion of HD. The authors use a dataset[44] of type 1 diabetes who had continuous glucose monitor data and applies LSTM techniques to learn a model. The paper suggests that training with the mean squared error (MSE) loss function is more effective than the negative loglikelihood (NLL) loss function for this dataset.

To ensure the confidentiality of HD, authors in [133] implemented an AT&T-based scheme for access control of medical records. The proposed scheme uses XACML for defining access policies. While storing, HD is encrypted using symmetric encryption. In [45] the authors describe several access control mechanisms and their applicability for ensuring the privacy of the HD. Discretionary Access Control (DAC) specifies per user per object based granular permissions which can be materialized using Access Control List (ACL) and Capability List (CL).

Sharing HD often requires confirmation that the HD has not been manipulated. Cryptographic signatures are generally used for this purpose. Additionally, it is also not required to share the complete HD, rather a partial HD is sufficient to describe the information that the doctors or the researchers need for their analysis. Content Extraction Signatures [122] are used for this purpose in the literature[78]. The work in [78] allows a patient to exclude certain information while sharing medical records with a doctor while retaining its authenticity. A three-layer architecture is proposed. In the first layer, EMR is created by data providers (e.g. doctors) and signed using the CES [121] scheme and then sent to the patients who own the EMR. The encrypted EMR is stored in the cloud and the indexes of the EMR are stored in a consortium blockchain. CES allows patients to exclude certain parts of the EMR while sharing.

### 4.3.2. Ledger based

Ensuring confidentiality is not the only concern to address, often immutability and traceability are also necessary. Hence a smartphone-based system is proposed in [152] in which the storage layer is maintained by a private blockchain cloud. The management layer works as a gateway that controls incoming and outgoing access. It also works as a database manager that stores heterogeneous personal data. The usage layer consists of entities that use HD for performing analysis. In the proposed system, a patient encrypts his/her healthcare data and shares the encrypted data along with the key while sending the data to a doctor. The doctor's system on the other hand destroys the replica of the data after a fixed period of time. In [38] the problem of storing HD originating from wearable devices is discussed. The data is stored on cloud storage and the hashes are stored on the blockchain. Blockchains are used not only for storing HD but also the HD-related transactions between doctors and patients [59], [130]. In [130] all transactions are performed using the Hyperledger Fabric blockchain. The proposed approach uses a smart contract (chaincode) that provides two functions, store and get, and all data is stored in the blockchain as key-value pairs. The throughput of the proposed system was found to be 50 transactions per second when implemented in Fabric version 1.0 with 4 peer nodes and one orderer node. In [59] the patients may delegate hospitals to encrypt their medical records and store them on semi-trusted cloud servers. The researchers consume the medical data from patients if the requirements are met. The data requirements are published via smart contracts. Patients who believe that their records meet the published requirements present zero-knowledge proof to the smart contract. Once qualified, the semi-honest cloud server transforms the encrypted medical data into an intermediate ciphertext that can be decrypted by the researcher. They evaluate the scheme in terms of three metrics (less computing cost, fewer startup nodes, and privacy protection). It has been found that the proposed scheme achieves the optimal performance with a 33% growth rate of block generation speed compared with the baseline. In comparison with PGHR, the proposed scheme offers faster proving and key generation and smaller key sizes while increasing verification time, which is still less than one second. In [90] a system is proposed in which the encrypted HD is stored in the cloud as a backup. A blockchain is used for maintaining and controlling data that is directly accessed by only one entity named the Private Accessible Unit. Users perform all transactions through this unit and get a Block ID in return. The block ID is used for retrieving healthcare records stored inside the block. The users must remember the block ID and authenticate with their credentials to access their data in the future. Computation time of generating cyphertext is considered as the effectiveness of the proposal. The results show that it depends linearly on the input size. The paper does not address the issue of interoperability between different entities (e.g. doctors, patients, institutions, etc.) and leaves that as future work. In [27] the proposed scheme the HD is encrypted with a symmetric key which is encrypted using the patient's public key. The digest and the hash of the medical record are signed and posted to the blockchain. The data can be decrypted by the patient and the patient may authorize a third-party agency to access that data. The authors have suggested the usage of Delegate Proof of Stake (DPOS) for consensus mechanisms in the blockchain network. In [145] and [67] also blockchain is used for access and sharing of HD. However, IPFS is used as a storage platform for storing the encrypted HD. To make HD searchable the authors in [156] propose a two blockchain-based system. Encrypted HD are stored in the private blockchain while the keywords are stored in a consortium blockchain accessible by multiple HIs. A token is generated, once a patient visits a doctor in a different HI, which permits the doctor to generate an HD and check past HDs if needed.

### 4.3.3. Learning based

Homomorphic encryption techniques are used for storage when the stored data has to be used for further analysis e.g. machine learning purposes. In [36] the authors have used homomorphic encryption to securely compute association rule mining while ensuring the privacy of the medical data. Different HIs containing medical data in the same schema perform local computation and use the result value as the secret while computing the Additive ElGamal ciphertext.

A Survey on Privacy of Health Data Lifecycle

A central server aggregates results from all institutions and sends the result back to the institutions where it is decrypted. The k-means algorithm is often used for analyzing HD. The usual scenario is that participants share their feature vector with an analyst who partitions the data into k different clusters based on a mutual similarity between two feature vectors. The authors conduct experiment using Wisconsin breast cancer dataset[141] and heart disease dataset[32]. The results suggest that the proposed approach reduces the computation time and the communication cost. In [94] a privacy-preserving k-NN classification technique is proposed. Instead of using Elgammal-based homomorphic encryption, the authors use Samir's Secret sharing and multi-party computation to find the top k similar records for any diagnosis query. In general, homomorphic cryptographical techniques support a limited number of mathematical operations, like addition or multiplication. But the fully homomorphic schemes support multiple operations which is very useful for scientific computation on HD. Experimental results suggest that the proposed work outperforms the baseline in terms of running time. In [125] the authors use the fully homomorphic scheme proposed in [37] to securely compute the average heart rate. The authors propose an architecture for a mobile healthcare network that collects encrypted body vitals from patients and perform three different computations, average heart rate, the long QT syndrome detection, and the chi-square tests using homomorphic encryption techniques.

The authors in [54] propose a mutual privacy-preserving k-means strategy based on homomorphic encryption. The proposed algorithm uses the Paillier cryptosystem while sharing the cluster center. To evaluate the effectiveness of the proposed strategy the algorithm is applied to three datasets. One is on public utilities in a city that consists of 240 latitudes and longitudes. The second one is Haberman's survival data set, which consists of the survival of 306 patients who had undergone surgery for breast cancer. The third one is the smartphone dataset for human activity recognition. It is found that the clustering results of the proposed algorithm are very close to the results of the original k-means algorithm. The reason behind the minor difference was the conversion of real numbers to integers. For the other two experiments, the results obtained are almost the same as the original k-means algorithm. t is observed that k-means is a very popular algorithm used for clustering HD.

In [23] the authors propose a framework for privacy-preserving big data analytics for healthcare data. In the proposed framework, first, the raw data are captured, then pre-processed for missing values and cleanup. The pre-processed data is then generalized for privacy preservation. Then that data is used for unsupervised learning. The framework is applied to a dataset of 1,79,625 HIV and TB patients from 1993 to 2014. The k-Means algorithm was used for unsupervised learning. With the experiments, the authors show that HD can be correlated with age groups and socio-economic backgrounds without hampering personal information. HD often includes graphical objects obtained from medical imaging procedures. Some image-specific techniques can be used instead of using generic cryptography for protecting such documents while enabling analysis.

### 4.3.4. Steganography based

In [149] Reversible data hiding techniques have been used for embedding data into medical images. Lesion regions of an image are identified and the contrast of that region is enhanced while embedding privacy data. The rest of the data is embedded into the non-lesion region using high-capacity embedding methods to achieve a higher payload. Auxiliary information about the lesion area and the embedding process is stored on the four sides of the images where there is no critical information. After embedding, Piecewise Linear Chaotic Map is used to generate a secret key which is used for the homomorphic encryption of that data. The authors use Shanon entropy as a metric of the randomness of ciphertext images, which is found to be close to 8 when conducting experiments using medical images e.g. MRI, CT scan, etc. It has also been found that the histogram of the encrypted images follows uniform distribution and there is no correlation between adjacent pixels in encrypted images. In [119], the authors propose a machine learning-based scheme for watermarking medical images against manipulations. A hash key is generated from the complete image using the SHA-256 algorithm after replacing the first LSB of each pixel with zero. The images are divided into two parts ROI (Region of Interest) and RONI (Region of no Interest). ROI is compressed using the LZW algorithm. The compressed ROI and the hash key is used for generating the fragile watermark which is inserted into RONI by replacing the first LSB bits of the pixels. A DNN framework is developed to extract the watermark data from the watermarked/attacked image. In summary, we see a consistent application of cryptographical access schemes which have been incorporated with blockchain-based solutions to ensure traceability and immutability. Recent works use cryptographic solutions for distributed storage (cloud, IPFS, etc..) solutions [67][145][116][27]. Homomorphic encryption is used to keep the data-at-rest compatible with machine learning systems, while watermarking schemes are used to protect medical imaging HD from adversarial manipulation.

### 4.4. Reconstruction Concerns

As mentioned in Section 3.5, HD can have different type of information. Even if the ID information is stripped out of the HD, an adversary can use the QI information and public information to identify a patient. In [51], a study has been presented based on social network posts related to rare diseases. The paper suggests that the person can be identified by the information provided.

#### 4.4.1. Anonymization based

Incorporating these techniques, a hybrid solution is proposed in [148] for preserving the privacy of medical data. In this solution, the original plain text data is partitioned into three tables. One table only with the medical information, and another table with the anonymized quasi-identifiable attributes. Additionally, an encrypted table is maintained that contains identifiers and quasi-identifiable attributes. The proposed framework consists of three components. The Data Merging component's responsibility is to merge the anonymized data with medical information when required. The plain text data if leaked is vulnerable to re-constructions by an adversary. The authors use their proposed approach on a dataset of 1 million medical records out of which 0.5 million are real world and the rest are generated by exchanging attributes. Global Certainty Penalty (GCP) is used as a metric to measure the quality of the anonymization. The experimental results suggest that the GCP increases linearly with the increase of the value of k. In [76] the authors propose a privacy-preserving data collection scheme using (a,k)-anonymization. Data is anonymized twice. Once on the client side and then again on the server side. In the proposed scheme clients submit anonymized data instead of submitting the original data. The authors use the Adult dataset[74] from UCI Machine Learning Repository. Experiments are conducted to evaluate the overhead of the proposed scheme. The computational complexity of the proposed scheme is found to be quadratic. In [26] the authors propose a cloudlet-based system for the storage and sharing of medical data. While transmitting sensor data from wearable devices to the cloudlet, encryption techniques have been used for protecting the privacy of the users. Similar patients (e.g. suffering from the same conditions) may share treatment information with each other. A trusted authority e.g. hospital calculates similarity and trust between the parties before sharing. While sharing, Identifiable and Quasi Identifiable data are encrypted while medical information is shared as plaintext. Collaborative Intrusion Detection System (IDS) is used to detect malicious intruders. The authors use a cloudlet mesh simulator to conduct experiments. The results suggest that the optimum configuration is to use 4 IDSs which obtain a 75% detection rate under a minimum system cost. In [53], the authors propose a KNN-based technique named Avatar to generate a synthetic dataset from a pseudo-anonymized dataset of the same size. Experiments have been conducted on AIDS and WBCD datasets and the results have been compared with the baselines (Synthpop and CT-GAN). The results suggest that the Hazard ratio values obtained using the synthetic AIDS dataset produced by Synthpop and Avatar are within the confidence interval of the original data, while CT-GAN induces underestimation. It has also been found on the WBCD dataset that, the F-scores obtained using synthetic dataset generated using Avatar are closest to the original. Some recent works focus on evaluating the privacy re-identification risks of anonymized data. In [25], the authors propose a decision tree-based approach to perform a qualitative analysis of anonymized HD. The authors in [68] propose a risk estimator based on the average of Gaussian copula and d-vine copula for estimating the re-identification risk of a dataset. The experimental results suggest that the proposed estimator outperforms other proposed estimators, such as the entropy estimator, Benedetti-Franconi estimator, and hypothesis test estimator in terms of estimation of the true risk of the dataset.

### 4.4.2. Hashing based

Often multiple HIs are interested in sharing HD among themselves to discover new knowledge about the patient by co-relating their HD. Such a process is called Linking. However, the institutions do not want to share the personal data of the patients to protect privacy. Hence, different privacy-preserving methods of Linking are proposed in the literature. While sharing an HD, encrypting the identifying attributes may be useful, when such encryption is performed in both databases located in two different institutions. Then the encrypted identifiers can be compared to find out the documents belonging to the same individuals. However, the slightest change in the identifiers may lead to a massive change in the cyphertext. Moreover, all HDs may not have the exact same set of identifiers. Therefore, the authors in [112] propose a protocol for privacy-preserving record linkage with encrypted identifiers allowing for errors in identifiers. The paper applies the Bloom filter on the q-grams of textual identifying data, such as a surname, using the Double Hashing scheme. To compare the similarity between two resulting bloom filters, a Dice coefficient is used. The authors use cryptographic hash functions, SHA1 and MD5 in their implementation. The authors compared the performance of the proposed method against the phonetic encoding-based unencrypted method on two German private administration databases. The results suggest that the performance of the proposed approach is similar to the performance of the unencrypted trigrams. In [19], Bloom Filter is combined with Cryptographic Long term Key (CLK). Multibit trees and Jaccard similarity are used for comparison. The authors experiment on patient databases of two Australian Hospitals. Results suggest that the combination of First and Last Name, DoB, and Sex outperforms other parameter sets. Due to legal restrictions, identifying data is often not shared. Hence, record linkage has to be performed by using the quasi-identifiable attributes of the HD. Moreover, there can also be randomly missing values in HD that may lead to false positive matches. Hence, in [131] the authors propose a lattice structure-based techniques to link HD with missing values. Hierarchical Classification Codes (e.g. ISCO-88) are often used in HD for classifying diseases. In such codes, the position of each character has an important significance. Authors in [113] use a Pseudorandom Number Generator (PRNG) to generate a random number based on the unigrams of these codes. The experimental results suggest that the proposed scheme outperforms standard and positional bloom filters in terms of Discriminatory power.

### 4.4.3. Multi party Computation based

Multi-party computation-based approaches are proposed in the literature for similar concerns. A secure deterministic protocol of data exchange using garbled circuits has been proposed in [24]. The proposed protocol matches medical records. Using garbled circuits. Both parties perform the same computation on a subset of data without sharing the actual medical records. One of the parties plays the role of generator that converts the function to be computed to a logic

| Paper | Implementation | Stages (Creation, Storage, Access, Sharing, Linking, Learning, Destruction) | Actors (Owner, Custodian, Borrower, Auditor) | Concerns | Techniques | Contributions |
|---|---|---|---|---|---|---|
| [42] | T | ☐■■☐☐☐☐ | ■ Patient / ■ Hospitals | Confidentiality | Samir's Secret Sharing | Secure access to HD that allow access by security teams in case of emergency. |
| [55] | X | ■☐☐☐☐☐☐ | | DoS, Eavesdropping | Zero Power Defence, RC5 | Harvests RF energy from external sources instead of using the main battery |
| [112] | D | ☐☐☐☐■☐☐ | | Reconstruction, Disclosure | q-gram, Double Hashing Bloom Filter | link HD from multiple HI using efficient calculation of similarity between cyphertext. |
| [103] | X | ■☐☐☐☐☐☐ | | Impersonation | Diffie Hellman | Security through proof of proximity. |
| [144] | X | ■☐☐☐☐☐☐ | | Eavesdropping | ECG based Key Agreement | Uses guardian device to jam transmission between implanted and malicious device |
| [129] | T | ☐■■■☐☐☐ | ■ Patient / ■ $HI^2$ / ■ Emergency | Confidentiality | CP-ABE | Associates different security level to HD and proposes threshold cryptosystem. |
| [72] | X | ■☐☐☐☐☐☐ | | Eavesdropping | AES | Proposes encryption module for secure transmission for wearable devices. |
| [123] | X | ■☐☐☐☐☐☐ | | Impersonation, DoS | MAC, Zero Power Defence | Harvesting RF energy to defend against DoS attacks. |
| [70] | X | ☐■■☐☐☐☐ | ■ Hospital | Confidentiality | RSA, Samir's Secret Sharing | Threshold cryptosystem, that distributes secret shares over the devices in the HI. |
| [118] | X | ■☐☐☐☐☐☐ | | Impersonation | RSS based Key Generation | Key generation using adaptive secret bit generation[96] technique. |
| [99] | D | ☐■■☐☐☐☐ | ■ Patient / ■ Hospital | Disclosure | Pseudoanonymity, TDE | Application of an analytical platform[98] for HD, developed at Houston Methodist Hospital. |
| [148] | X | ☐■■■☐☐☐ | ■ Patient / ■ ■ Hospital / ■ RHCP[!] / ■ $RI^2$ | Confidentiality, Reconstruction, Re-Identification | Vertical Partitioning, K-Anonymity, T-closeness, AES, RSA, Hybrid Search | HD is stored into multiple tables. Some plaintext, anonymized, encrypted. |
| [143] | X | ☐☐☐☐☐☐☐ | | Disclosure, Accessibility | Distributed Randomization, K-anonymity | Algorithm to combine K-anonymity and Distributed Randomization |
| [49] | X | ☐☐☐☐■☐☐ | | Disclosure | k-means | Proposes privacy preserving non-cryptographic clustering technique. |
| [152] | X | ☐■☐☐☐■■ | ■■ Patient / ■ Blockchain / ■ Doctor | Confidentiality, Manipulation, Accessibility | Blockchain, Symmetric | Architecture of Healthcare Data Gateway and mobile application to incorporate blockchain based storage and access. |
| [28] | X | ☐☐☐☐■☐☐ | | Disclosure | Bloom Filter, k-NN | Designs a framework for record linkage to deal with missing values in HD. |
| [19] | X | ☐☐☐☐■☐☐ | | Reconstruction, Disclosure | Cryptographic Longterm Key, Bloom Filter | Compares CLK based record linkage with clear-text probabilistic record linkage. |
| [84] | X | ☐☐☐☐■☐☐ | | Reconstruction, Disclosure | MPSI, Additive ElGamal, Bloom Filter | Privacy preserving analysis of HD distributed over multiple institutions. |
| [125] | X | ☐■☐☐☐■☐ | ■ Patient | Confidentiality, Accessibility | Fully Homomorphic Encryption | Mobile healthcare network for long QT syndrome detection. |
| [124] | X | ☐☐■☐☐☐☐ | | Confidentiality | AES, IBE | Proposes a system of HD sharing using symmetric and asymmetric encryption. |
| [27] | X | ☐■■☐☐☐☐ | ■ Patient | Confidentiality | Blockchain | Proposes a blockchain based solution for storage and sharing of HD. |
| [24] | X | ☐☐☐☐■☐☐ | | Disclosure | Garbled Circuit | Approximate matching mechanism to link HD spread across different institutions. |
| [94] | X | ☐☐☐☐■☐☐ | ■ ■ Hospital | Disclosure | k-NN, Shamir's secret sharing | Matching HD with a set of symptoms through collaboration of multiple HI. |
| [78] | X | ☐■■■☐☐☐ | ■ Patient / ■ Blockchain / ■ Doctor / ■ Consortium | SPoF, Confidentiality, Manipulation | Pseudoanynymity, Blockchain, CP-ABE, CES[121] | Proposes secure privacy preserving system with off-chain cloud storage for HD while storing indexes in a consortium blockchain. |
| [15] | X | ■☐☐☐☐☐☐ | | Eavesdropping, Manipulation | Henon Scheme, Chaotic Map | Uses signature algorithm to protect implanted devices from MiTM attacks. |
| [76] | X | ■☐☐☐☐☐☐ | | Re-Identification | Vertical Partitioning, (a,k)-anonymity | Anonymized HD transmission from IoT devices. |
| [156] | X | ☐■■☐☐☐☐ | ■ Patient / ■ Hospital / ■ Consortium | Confidentiality, Manipulation, Accessibility | Bilinear Mapping, Blockchain, Smart Contract, PEKS[16] | Proposes private blockchain for storing HD and consortium blockchain for storing keywords. |
| [137] | X | ■☐☐☐☐☐☐ | | Eveasdropping, Impersonation, DoS | Physical Unclonable Function | non-cryptographic solution for secure communication with BAN sensors. |
| [35] | X | ☐■☐☐■☐☐ | ■ ■ $HI^{14}$ | Disclosure, Re-Identification | Vertical Partitioning, RSA, Blind Signature, Association Rule Mining | Privacy preserving analysis of genome information. |

| Paper | Implementation | Stages (Creation, Storage, Access, Sharing, Linking, Learning, Destruction) | Actors (Owner, Custodian, Borrower, Auditor) | Concerns | Techniques | Contributions |
|---|---|---|---|---|---|---|
| [38] | X | ■■□□□□□ | Patient (Owner); Cloud (Custodian); HI[14] (Borrower) | SPoF, Confidentiality, Disclosure, Manipulation | Blockchain, Smart Contract; Deffie Hellman; Ring Signature; ARX, Public Key Cryptography | Proposes a smart contract based system to analyze and store HD in the cloud while sending alerts back to the patients in an IoT environment. |
| [149] | X | □■□□□□□ | | Confidentiality, Accessibility | Reversible Data Hiding | Framework for embedding privacy data into medical imaging HD. |
| [90] | X | □■■□□□□ | Patient (Owner); Cloud (Custodian) | Confidentiality, Disclosure, Manipulation | Pseudonymity; Blockchain; Elliptic Curve Cryptography | Proposes information storage and retrieval system for healthcare where encrypted HD is stored in the blockchain. |
| [145] | X | ■■■■□□□ | Patient (Owner); Hospital (Custodian); Doctor (Borrower) | Confidentiality, Disclosure, Manipulation | Pseudonymity; Blockchain; IPFS, AES | Uses multiple blockchains for IoT data and diagnosis where encrypted HD is stored in IPFS. |
| [67] | X | □■■■□□□ | Patient (Owner); Hospital (Custodian); Doctor (Borrower) | Confidentiality, Disclosure, Manipulation | Blockchain; IPFS; AES | Encrypted HD is stored in IPFS while blockchain is used to store the hashes. |
| [113] | X | □□□□■□□ | | Disclosure | Hierarchy Preserving Bloom Filter | Introduces new encoding techniques for Hierarchical codes used in HD |
| [47] | X | □□□□□■□ | | Confidentiality, Disclosure, Manipulation | Blockchain; Smart Contract; Pallier | Statistical analysis on HD stored in a distributed blockchain network. |
| [54] | X | □□□□□■□ | | Disclosure, Accessibility | Paillier; K-Means | Clustering of HD while securely involving a data analyst and third party cloud. |
| [101] | X | □□■□□□□ | Patient (Owner); Emergency (Borrower) | Confidentiality, Accessibility | Blockchain; Smart Contract | Proposes blockchain based scheme to enable access to HD in emergency situations. |
| [11] | X | ■□□□□□□ | | Eveasdropping | ECG based Key Agreement; QRS complex detection, LFSR | Proposes ECG based Key agreement scheme based on QRS complex detection |
| [157] | X | □□□□□■□ | Patient (Owner); Cloud (Borrower) | Accessibility | Decision Tree; kNN | Proposes decision tree evaluation scheme for medical diagnosis. |
| [26] | X | ■■■■□□□ | Patient (Owner, Borrower); Hospital (Custodian) | Confidentiality, Accessibility | Vertical Partitioning; NTRU, Additive Homomorphic Encryption | Proposes collaborative Intrution detection system based on cloudlet mesh. |
| [46] | X | □□□□□■■ | Patient (Owner) | Confidentiality, Traceability | LSTM; KP-ABE, Markle Hash Tree | Raise alarm by monitoring HD originating from wireless wearable devices. |
| [147] | X | ■□□□□□□ | | Impersonation | Autoregression; LSTM | RSSI based predictive model to distinguish malicious frames from legitimate ones over BAN. |
| [100] | X | □■■□□□□ | Patient (Owner); Doctor (Borrower) | Confidentiality | Fragmentation; AES | Small subset of fragmented HD is encrypted and stored in user's smartphone while the rest is stored in the cloud. |
| [130] | X | □■□■□□□ | Patient (Owner); Consortium (Custodian, Auditor); Doctor, Pharmacist, Insurer (Borrower) | Confidentiality, Accessibility, Manipulation | Diffie Hellman; Smart Contract; SIFF[160] | Proposes blockchain based storage of HD accessed through smart contracts aided by QR codes. |
| [59] | X | □■□■□□■ | Patient (Owner); Hospital (Custodian); RI[2] (Borrower) | Confidentiality, SPoF, Traceability | Smart Contract; zk-SNARK | Proposes a framework for the RI[2] to publish the requirements and the patients prove that their HD satisfies that. |
| [23] | X | □□□□□■□ | | Accessibility | Vertical Partitioning; K-Means | Proposes a privacy aware big data analytics framework for HD. |
| [110] | X | ■□□□□□□ | | Eavesdropping | Locality Sensitive Hashing; HMAC | Transmits irreversible signature of HD to defend man in the middle attacks. |
| [131] | X | □□□□■□□ | | Disclosure | Lattice based Bloom Filter | HD linking technique with missing data. |
| [68] | X | □□□■□□□ | | Re-Identification | Copula Method | Proposes metrics for assesing privacy risk of anonymized HD |
| [36] | X | □□□□□■□ | Patient (Owner); Hospital (Custodian, Borrower) | Accessibility, Disclosure | Association Rule Mining; Additive ElGamal | Aggregate functions over horizontally partitioned data using a central server. |
| [53] | X | □□□■□□□ | | Re-Identification | KNN; Avatar | Generate anonymized synthetic dataset from original dataset. |
| [119] | X | □■■□□□□ | | Manipulation | DNN, LZW | Watermarking solution for medical HD. |

[1] Regional Healthcare Collaboration Platform    [2] Research Institute

Table 4: Comprehensive overview of the state-of-the-art techniques addressing different HD privacy concerns involving different actors interacting with HD at different stages of HD lifecycle. (T,X,D denotes, Theoretical, Experimental and Deployed respectively.)

circuit and creates a garbled input that corresponds to the actual input. It is not possible to get back the original input from the garbled input. However, the other party that works as an evaluator can compute the output of the garbled circuit using the garbled input that it receives. The paper implements the proposed solution and evaluates its performance. In [84] the authors propose a scheme for multi-party private set intersection, based on bloom filter and *(n, n)-threshold exElGamal* encryption under the honest-but-curious model. The protocol uses joint decryption of an (n, n)-threshold exElGamal among *n* players. The results show that the proposed protocol is faster than the baseline.

In summary, we observe that various anonymization-based solutions have been proposed to address reconstruction /re-identification concerns. In recent work, not only tabular data but also imaging data is considered. Moreover, recent works have also focused on proposing a metric for assessing privacy risk from anonymized data. Hashing technique such as the bloom filter has been widely used for multi-party computations such as set intersection and record linkage. However, different variations of bloom filters are being applied along with other hashing and cryptographic schemes.

## 5. Conclusion, Challenges and Future Directions

In this paper, we provide a comprehensive review of HD privacy concerns and applicable techniques to address these concerns. Specifically, we identify different stages that HD goes through in its lifecycle, along with different privacy risks imposed on HD at each stage. We review a number of techniques that can be used to reduce the privacy risks of HD, ranging from cryptography and hashing to ledger techniques. Moreover, we propose a taxonomy that triangulates between lifecycle stages, privacy concerns, and privacy protection techniques, and as such allows us to identify exactly which techniques are applicable to particular privacy concerns at a particular HD lifecycle stage.

In summary, our review study reveals that implantable and wearable devices are resource constrained and vulnerable to network-based attacks. Moreover, the application of computationally intensive cryptographic operation may lead to energy drains [137], which can be fatal in the case of implanted body sensors. Network concerns are focused on impersonation, spoofing, and Man in the Middle attacks. Different technical works consider different adversary models. Solutions span from cryptographic communication schemes to monitoring the traffic over the wireless network. Some works use environmental or biological phenomena like RSS or ECG to calculate the proximity of nodes and detect adversaries. Some works apply machine learning-based techniques to detect malicious behaviors. However, in spite of several academic solutions proposed, the devices used in the healthcare industry suffer from security risks [139]. In most cases, backups of imaging HD are stored as plain text and do not implement a validation mechanism, which is vulnerable to confidentiality, eavesdropping, and manipulation concerns [139].

Cryptographic techniques are heavily used to address storage concerns. Role-based and Attribute-based access policies are often used for ensuring privacy on HD in the custody of HIs. In order to ensure the integrity of HD, cryptographic signatures are often used. However, Content Extractable Signatures (CES) [121] are used instead of conventional signatures when a partial HD is shared. Not only that but also the actions performed on HD by other entities involved in its lifecycle have to be documented and secured from manipulations. In order to ensure immutability, distributed ledger techniques are often used to document HD transactions. Some researchers even propose storing HD in blockchain to mitigate Single point of failure concerns originating from centralized storage solutions. However, in many works, HD is stored off-chain on the cloud and their hashes are stored in the blockchain. The wide adoption of encryption techniques makes it very difficult to perform knowledge discovery from the existing data. Hence, researchers often use homomorphic encryption to perform computation on the cyphertexts. However, Partial homomorphic encryption can only support a subset of operations while a fully homomorphic encryption algorithm is computationally expensive. Recent advancements in the integrations of machine learning techniques with fully homomorphic encryption are making such approaches more and more practical [142]. Moreover, performing knowledge discovery on encrypted HD is not risk-free, because the homomorphically encrypted ciphertexts can be used for mathematical operations without decryption. Thus, it is also essential to secure access to the ciphertext [89]. The application of cryptography increases the complications of the HD management system. Hence, it is essential to maintain the tradeoff between cryptographic strength and maintainability.

In order to ensure the confidentiality of HD even when the storage system is compromised, HD is often anonymized while being stored. HD is also partitioned vertically, the identified parts of it are encrypted and the sensitive medical information is anonymized. In that case, the attacker can not get access to sensitive information even if it gets access to the storage system. Such anonymization techniques are also used while sharing HD between multiple institutions which intend to discover new knowledge by linking multiple HDs belonging to the same patient. Hashing-based techniques are applied to the identifying information to transform them into comparable but non-reversible values which can be used to link related HD without sacrificing patients' privacy. The application of hashing techniques is not only limited to identifying information about the patients. Even medical information like disease classifications is also hashed for comparison without disclosure. The process of distributed knowledge discovery may require the application of functions on the subset of data and exchanging output. However, in order to ensure the correctness of such computations, secure multiparty computation protocols are used.

Moreover, a recent study using textual analysis on social networks reveals that there is a clear patient demand for more control over HD, in terms of transparency, access, and interoperability[107]. To address the ethical, regulatory, safety, and quality concerns, the authors in [105] have proposed a four-component governance model. The components are fairness, transparency, trustworthiness, and accountability.

Finally, we conclude by discussing challenges and open research directions in the development of approaches for improving the privacy of HD in its lifecycle.

### 5.1. Emergency Access of HD

We have found works on allowing un-consented access to HD in emergency situations when the patient cannot perform cryptographic actions (e.g. unconscious). Some works propose threshold cryptosystem-based techniques that ensure privacy through the collaboration of multiple entities[42][129]. The other blockchain-based techniques[101] emphasize the immutability of information regarding the emergency access event while using consensus for approval. Integration of these two techniques may provide secure and privacy-aware access to HD in case of emergency. However, what happens to that HD after the emergency period is over is also important. Verifiable destruction techniques can ensure that once the emergency period is over, HD can no longer be accessed by the emergency staff or doctors.

### 5.2. Access and Sharing of Partial HD

The custodian or the borrower might not need the complete HD for the purpose of treatment, or the patient might not be willing to share all parts of his/her HD for privacy concerns. There have been few works[78] based on CES[121] to address this issue. However, more research is required to incorporate such techniques in critical healthcare situations, such as emergency access. Moreover, medical research may also facilitate such techniques by giving the patients option to participate with verifiable partial HD. Such practices may encourage people's participation in knowledge discovery.

### 5.3. Usability and Cryptography

Moreover, the encrypted data makes it more difficult to use machine learning applications. Although homomorphic encryption has been used in the literature for similar use cases, some algorithms require the encryption schemes to be fully homomorphic, which is very slow[142]. Hence, new advancements are required in the field of fully homomorphic encryption to make privacy-preserving HD analysis easier. Additionally, the homomorphically encrypted ciphertexts can be used for mathematical operations without decryption. Hence it is also essential to secure access to the ciphertext [89].

### 5.4. Interoperability of Medical Devices

Different medical devices use different operating systems and platforms. This heterogeneity makes it difficult to assess the security risks and provide a generic solution for all devices [86]. This also makes it difficult to develop software, in terms of interoperability [89]. Moreover, there is a lack of standardization in the communication protocols used for operations on medical devices. Although there exists standardization of guidelines and good practices [63][65][62][66][64] these studies do not deal with the fundamental S&P issues [89].

### 5.5. Traceable Anonymization

In [89] the authors considered the tradeoff between anonymization and traceability. Complete anonymization makes HD untraceable, while some conditional anonymization schemes provide conditional identity disclosure under special circumstances while making the data more traceable. There is a tradeoff between privacy and traceability and it is an open challenge to provide a strong privacy guarantee while keeping the data traceable.

### 5.6. Verifiable Destruction

Destruction is the last stage of the HD lifecycle. The privacy-centric goal of this stage is to ascertain that the HD is irreversibly destroyed. Such destruction ensures that even if the storage gets compromised the adversary cannot gain any knowledge about the HD. Although there is a lot of research work around the technicalities of secure data deletion, we have not found many works that target a healthcare scenario. The work in [54] targets HD originating from wearable devices and uses a Markle Tree-based technique to establish verifiable deletion of data. However, there can be many different types of HD associated with different contexts and deletion can be requested by the owners at any stage of the HD lifetime. Future research on HD privacy may address these challenges.

### Acknowledgements

This work is supported by the Research Council of Norway, grant number 288106.

A Survey on Privacy of Health Data Lifecycle# References

[1] , 1956. The Indian Medical Council Act, 1956 (102 of 1956). URL: http://www.bareactslive.com/ACA/ACT725.HTM.

[2] , 2002. The Indian Medical Council (Professional Conduct, Etiquette And Ethics) Regulations, 2002. URL: https://indiankanoon.org/doc/100527417/.

[3] , 2016. Regulation (eu) 2016/679 of the european parliament and of the council of 27 april 2016 on the protection of natural persons with regard to the processing of personal data and on the free movement of such data, and repealing directive 95/46/ec (general data protection regulation). Official Journal of the European Union L119.

[4] , 2021. A Survey on Healthcare Data: A Security Perspective. ACM Transactions on Multimedia Computing, Comm. and Applications 17.

[5] , 2021. Study finds issues with electronic ankle monitors used as alternative to incarceration. URL: https://gwtoday.gwu.edu/study-finds-issues-electronic-ankle-monitors-used-alternative-incarceration.

[6] Abouelmehdi, K., Beni-Hessane, A., Khaloufi, H., 2018. Big healthcare data: preserving security and privacy. Journal of Big Data 5.

[7] Alexander, M., 2018. The newest jim crow recent criminal justice reforms contain the seeds of a frightening system of "e-carceration.". URL: https://www.nytimes.com/2018/11/08/opinion/sunday/criminal-justice-reforms-race-technology.html.

[8] Altawy, R., Youssef, A.M.R.M., Member, S., 2016. Security Tradeoffs in Cyber Physical Systems : A Case Study Survey on Implantable Medical Devices. IEEE Access 4.

[9] Arbabi, M.S., Lal, C., Veeraragavan, N.R., Marijan, D., Nygård, J.F., Vitenberg, R., 2023. A survey on blockchain for healthcare: Challenges, benefits, and future directions. IEEE Communications Surveys & Tutorials 25.

[10] Aziz, M.M.A., Sadat, M.N., Alhadidi, D., Wang, S., Jiang, X., Brown, C.L., Mohammed, N., 2017. Privacy-preserving techniques of genomic data-a survey. Briefings in Bioinformatics 20.

[11] Bai, T., Lin, J., Li, G., Wang, H., Ran, P., Li, Z., Li, D., Pang, Y., Wu, W., Jeon, G., 2019. A lightweight method of data encryption in BANs using electrocardiogram signal. Future Generation Computer Systems 92.

[12] Banerjee, S.S., Hemphill, T., Longstreet, P., 2018. Wearable devices and healthcare: Data sharing and privacy. Information Society 34.

[13] Barcena, M., Wueest, C., Lau, H., 2014. How safe is your quantified self? Symantec URL: https://www.symantec.com/content/dam/symantec/docs/white-papers/how-safe-is-your-quantified-self.pdf.

[14] Bartoletti, I., 2019. Ai in healthcare: Ethical and privacy challenges, in: Riaño, D., Wilk, S., ten Teije, A. (Eds.), Artificial Intelligence in Medicine, Springer International Publishing, Cham.

[15] Belkhouja, T., Mohamed, A., Al-Ali, A.K., Du, X., Guizani, M., 2018. Light-Weight Solution to Defend Implantable Medical Devices against Man-In-The-Middle Attack. 2018 IEEE Global Communications Conference, GLOBECOM 2018 - Proceedings .

[16] Boneh, D., Crescenzo, G.D., Ostrovsky, R., Persiano, G., 2004. Public key encryption with keyword search, in: Cachin, C., Camenisch, J. (Eds.), Advances in Cryptology - EUROCRYPT 2004, International Conference on the Theory and Applications of Cryptographic Techniques, Interlaken, Switzerland, May 2-6, 2004, Proceedings, Springer.

[17] Bonomi, L., Huang, Y., Ohno-Machado, L., 2020. Privacy challenges and research opportunities for genomic data sharing. Nature Genetics 52.

[18] Briganti, G., Le Moine, O., 2020. Artificial intelligence in medicine: Today and tomorrow. Frontiers in Medicine 7. URL: https://www.frontiersin.org/article/10.3389/fmed.2020.00027.

[19] Brown, A.P., Borgs, C., Randall, S.M., Schnell, R., 2017. Evaluating privacy-preserving record linkage using cryptographic long-term keys and multibit trees on large medical datasets. BMC Medical Informatics and Decision Making 17.

[20] Burleson, W., Carrara, S., 2013. Security and privacy for implantable medical devices. Security and Privacy for Implantable Medical Devices 9781461416.

[21] Burns, A.J., Johnson, M.E., Honeyman, P., 2016. A brief chronology of medical device security. Communications of the ACM 59.

[22] Camara, C., Peris-Lopez, P., Tapiador, J.E., 2015. Security and privacy issues in implantable medical devices: A comprehensive survey. Journal of Biomedical Informatics 55.

[23] Chauhan, R., Kaur, H., Chang, V., 2021. An Optimized Integrated Framework of Big Data Analytics Managing Security and Privacy in Healthcare Data. Wireless Personal Communications 117.

[24] Chen, F., Jiang, X., Wang, S., Schilling, L.M., Meeker, D., Ong, T., Matheny, M.E., Doctor, J.N., Ohno-Machado, L.e.a., 2018a. Perfectly Secure and Efficient Two-Party Electronic-Health-Record Linkage. IEEE Internet Computing 22.

[25] Chen, M., Cang, L.S., Chang, Z., Iqbal, M., Almakhles, D., 2023. Data anonymization evaluation against re-identification attacks in edge storage. Wireless Networks 6.

[26] Chen, M., Qian, Y., Chen, J., Hwang, K., Mao, S., Hu, L., 2020. Privacy Protection and Intrusion Avoidance for Cloudlet-Based Medical Data Sharing. IEEE Transactions on Cloud Computing 8.

[27] Chen, Y., Ding, S., Xu, Z., Zheng, H., Yang, S., 2018b. Blockchain-Based Medical Records Secure Storage and Medical Service Framework. Journal of Medical Systems 43.

[28] Chi, Y., Hong, J., Jurek, A., Liu, W., O'Reilly, D., 2017. Privacy preserving record linkage in the presence of missing values. Information Systems 71.

[29] Commission, I.E., for Standardization, I.O., 2011. 6-21- ISO 29100: Information technology — Security techniques — Privacy framework Technologies 2011.

[30] Coventry, L., Branley, D., 2018. Cybersecurity in healthcare: A narrative review of trends, threats and ways forward. Maturitas 113.

[31] Defend, B., Salajegheh, M., Fu, K., Inoue, S., 2008. Protecting Global Medical Telemetry Infrastructure Executive Summary .

[32] Detrano, R., Janosi, A., Steinbrunn, W., Pfisterer, M., Schmid, J.J., Sandhu, Sarbjit, e.a., 1989. International application of a new probability algorithm for the diagnosis of coronary artery disease. The American journal of cardiology 64.

[33] Deuber, D., Christin, N., 2022. Assessing Anonymity Techniques Employed in German Court Decisions : A De-Anonymization Experiment. Proceedings of the 31st USENIX Security Symposium, Security 2022 .

[34] Dolan, B., 2014. In-depth: Consumer health and data privacy issues beyond hipaa. URL: https://www.mobihealthnews.com/33393/in-depth-consumer-health-and-data-privacy-issues-beyond-hipaa.

[35] Domadiya, N., Rao, U.P., 2019. Privacy Preserving Distributed Association Rule Mining Approach on Vertically Partitioned Healthcare Data. Procedia Computer Science 148.

[36] Domadiya, N., Rao, U.P., 2022. ElGamal Homomorphic Encryption-Based Privacy Preserving Association Rule Mining on Horizontally Partitioned Healthcare Data. Journal of The Institution of Engineers (India): Series B .

[37] Dowlin, N., Gilad-Bachrach, R., Laine, K., Lauter, K., Naehrig, M., Wernsing, J., 2017. Manual for Using Homomorphic Encryption for Bioinformatics. Proceedings of the IEEE 105.

[38] Dwivedi, A.D., Srivastava, G., Dhar, S., Singh, R., 2019. A decentralized privacy-preserving healthcare blockchain for IoT. Sensors (Switzerland) 19.

[39] Dyer, C., 2013. Police investigate allegations of data falsification at Essex hospital. BMJ (Clinical research ed.) 347.

[40] El Majdoubi, D., El Bakkali, H., Sadki, S., Maqour, Z., Leghmid, A., 2022. The Systematic Literature Review of Privacy-Preserving Solutions in Smart Healthcare Environment. Security and Communication Networks 2022.

[41] Emam, K.E., Kosseim, P., 2009. Privacy interests in prescription data, part 2: Patient privacy. IEEE Security and Privacy 7.

[42] Eskeland, S., Oleshchuk, V.A., 2007. EPR access authorization of medical teams based on patient consent P-118.
Bose et al.: *Preprint submitted to Elsevier*  Page 22 of 26

[43] Forensics, E., 2021. Electronic medical records manipulated post lawsuit. URL: https://enigmaforensics.com/blog/electronic_medical_records_manipulation/.

[44] Fox, I., Ang, L., Jaiswal, M., Pop-Busui, R., Wiens, J., 2018. Deep multi-output forecasting: Learning to accurately predict blood glucose trajectories, in: ACM SIGKDD International Conference on Knowledge Discovery & Data Mining.

[45] Gajanayake, R., Iannella, R., Sahama, T., 2014. Privacy oriented access control for electronic health records. Electronic Journal of Health Informatics 8.

[46] Ge, C., Yin, C., Liu, Z., Fang, L., Zhu, J., Ling, H., 2020. A privacy preserve big data analysis system for wearable wireless sensor network. Computers and Security 96.

[47] Ghadamyari, M., Samet, S., 2019. Privacy-Preserving Statistical Analysis of Health Data Using Paillier Homomorphic Encryption and Permissioned Blockchain. Proceedings - 2019 IEEE International Conference on Big Data, Big Data 2019 .

[48] Ghazarian, A., Zheng, J., El-Askary, H., Chu, H., Fu, G., Rakovski, C., 2021. Increased Risks of Re-identification for Patients Posed by Deep Learning-Based ECG Identification Algorithms. Proceedings of the Annual International Conference of the IEEE Engineering in Medicine and Biology Society, EMBS .

[49] Gheid, Z., Challal, Y., 2016. Efficient and privacy-preserving k-means clustering for big data mining. IEEE TrustCom .

[50] Golle, P., 2006. Revisiting the Uniqueness of Simple Demographics .

[51] Gow, J., Moffatt, C., Blackport, J., 2020. Participation in patient support forums may put rare disease patient data at risk of re-identification. Orphanet Journal of Rare Diseases 15.

[52] Greene, T., 2022. Nhs trust 'deliberately' deleted up to 90,000 emails before tribunal hearing. URL: https://www.computerweekly.com/news/252522787/NHS-trust-deliberately-deleted-up-to-90000-emails-before-tribunal-hearing.

[53] Guillaudeux, M., Rousseau, O., Petot, J., Bennis, Z., Dein, C.A., Goronflot, T., Vince, N., Limou, S., Karakachoff, M., Wargny, M., et al., 2023. Patient-centric synthetic data generation, no reason to risk re-identification in biomedical data analysis. NPJ Digital Medicine 6.

[54] Guo, X., Lin, H., Wu, Y., Peng, M., 2020. A new data clustering strategy for enhancing mutual privacy in healthcare IoT systems. Future Generation Computer Systems 113.

[55] Halperin, D., Clark, S.S., Fu, K., Heydt-Benjamin, T.S., Defend, B., Kohno, T., et al., 2008. Pacemakers and implantable cardiac defibrillators: Software radio attacks and zero-power defenses. Proceedings - IEEE Symposium on Security and Privacy .

[56] Hannon, B.P.d., 2021. Kentucky woman, 59, with aggressive breast cancer sues hospital after worker sent her the wrong letter saying she was in the clear and then 'covered it up by editing electronic records'. URL: https://www.dailymail.co.uk/news/article-9373713/Kentucky-woman-breast-cancer-sues-hospital-WRONG-letter-says-clear.html.

[57] Hathaliya, J.J., Tanwar, S., 2020. An exhaustive survey on security and privacy issues in Healthcare 4.0. Computer Communications 153.

[58] Henriksen-bulmer, J., Jeary, S., 2016. Re-identification attacks—A systematic literature review Jane. International Journal of Information Management 36.

[59] Huang, H., Zhu, P., Xiao, F., Sun, X., Huang, Q., 2020. A blockchain-based scheme for privacy-preserving and secure sharing of medical data. Computers and Security 99.

[60] Isidore, C., 2015a. Apple, at & t fight sale of radioshack customer data. URL: https://money.cnn.com/2015/05/15/news/companies/radioshack-apple-att/index.html.

[61] Isidore, C., 2015b. Radioshack trying to sell data of 100 million customers. URL: https://money.cnn.com/2015/03/25/news/companies/radioshack-customer-data/?iid=EL.

[62] ISO, 2006. IEC 62304:2006 Medical device software — Software life cycle processes. URL: https://www.iso.org/standard/38421.html.

[63] ISO, 2012. ISO/IEC 27032:2012 Information technology — Security techniques — Guidelines for cybersecurity. URL: https://www.iso.org/standard/44375.html.

[64] ISO, 2016. IEC 82304-1:2016 Health software — Part 1: General requirements for product safety. URL: https://www.iso.org/standard/59543.html.

[65] ISO, 2017. ISO/TR 80002-2:2017 Medical device software — Part 2: Validation of software for medical device quality systems. URL: https://www.iso.org/standard/60044.html.

[66] ISO, 2021. IEC 80001-1:2021 Application of risk management for IT-networks incorporating medical devices — Part 1: Safety, effectiveness and security in the implementation and use of connected medical devices or connected health software. URL: https://www.iso.org/standard/72026.html.

[67] Jayabalan, J., Jeyanthi, N., 2022. Scalable blockchain model using off-chain IPFS storage for healthcare data security and privacy. Journal of Parallel and Distributed Computing 164.

[68] Jiang, Y., Mosquera, L., Jiang, B., Kong, L., El Emam, K., 2022. Measuring re-identification risk using a synthetic estimator to enable data sharing. PLoS ONE 17.

[69] Jillson, E., Plant, M., 2021. Developer of popular women's fertility-tracking app settles ftc allegations that it misled consumers about the disclosure of their health data. URL: https://www.ftc.gov/news-events/news/press-releases/2021/01/developer-popular-womens-fertility-tracking-app-settles-ftc-allegations-it-misled-consumers-about.

[70] Jose, J.T., Anju, S., 2013. Threshold Cryptography Based Secure Access Control for Electronic Medical Record in an Intensive Care Unit 2.

[71] Keshta, I., Odeh, A., 2021. Security and privacy of electronic health records: Concerns and challenges. Egyptian Informatics Journal 22.

[72] Kim, J., Lee, B.J., Yoo, S.K., 2013. Design of real-time encryption module for secure data protection of wearable healthcare devices. Proceedings of the Annual International Conference of the IEEE Engineering in Medicine and Biology Society, EMBS .

[73] Kissner, L., Song, D., 2005. Privacy-preserving set operations, in: Annual International Cryptology Conference, Springer.

[74] Kohavi, R., et al., 1996. Scaling up the accuracy of naive-bayes classifiers: A decision-tree hybrid., in: Kdd.

[75] Li, C., Raghunathan, A., Jha, N.K., 2011. Hijacking an insulin pump: Security attacks and defenses for a diabetes therapy system. 2011 IEEE 13th International Conference on e-Health Networking, Applications and Services, HEALTHCOM 2011 .

[76] Li, H., Guo, F., Zhang, W., Wang, J., Xing, J., 2018. (a,k)-Anonymous Scheme for Privacy-Preserving Data Collection in IoT-based Healthcare Services Systems. Journal of Medical Systems 42.

[77] Li, N., 2011. t-Closeness : Privacy Beyond k-Anonymity and l-Diversity .

[78] Liu, J., Li, X., Ye, L., Zhang, H., Du, X., Guizani, M., 2018. Bpds: A blockchain based privacy-preserving data sharing for electronic medical records, in: 2018 IEEE Global Communications Conference (GLOBECOM), pp. 1–6.

[79] Machanavajjhala, A., Gehrke, J., Kifer, D., Venkitasubramaniam, M., 2006. L-diversity: privacy beyond k-anonymity, in: 22nd International Conference on Data Engineering (ICDE'06), IEEE.

[80] MacRae, J., Dobbie, S., Ranchhod, D., 2012. Assessing re-identification risk of de-identified health data in new zealand. Health Care and Informatics Review Online 16.

[81] Malina., L., Hajny., J., Dzurenda., P., Ricci., S., 2018. Lightweight ring signatures for decentralized privacy-preserving transactions, in: Proceedings of the 15th International Joint Conference on e-Business and Telecommunications - SECRYPT, INSTICC. SciTePress.

[82] Marijan, D., Lal, C., 2022. Blockchain verification and validation: Techniques, challenges, and research directions. Computer Science Review 45, 100492. URL: https://www.sciencedirect.com/science/article/pii/S1574013722000314, doi:https://doi.org/10.1016/j.cosrev.2022.100492.

A Survey on Privacy of Health Data Lifecycle[83] Mirsky, Y., Mahler, T., Shelef, I., Elovici, Y., 2019. CT-GAN: malicious tampering of 3d medical imagery using deep learning, in: 28th USENIX Security Symposium, USENIX Security 2019, Santa Clara, CA, USA, August 14-16, 2019, USENIX Association.

[84] Miyaji, A., Nakasho, K., Nishida, S., 2017. Privacy-Preserving Integration of Medical Data: A Practical Multiparty Private Set Intersection. Journal of Medical Systems 41.

[85] Munro, D., 2016. Data breaches in healthcare totaled over 112 million records in 2015. URL: https://www.forbes.com/sites/danmunro/2015/12/31/data-breaches-in-healthcare-total-over-112-million-records-in-2015/?sh=4e331bf67b07.

[86] Newaz, A.I., Sikder, A.K., Rahman, M.A., Uluagac, A.S., 2021. A Survey on Security and Privacy Issues in Modern Healthcare Systems. ACM Transactions on Computing for Healthcare 2.

[87] Nguyen, D.C., Pham, Q.V., Pathirana, P.N., Ding, M., Seneviratne, A., Lin, Z., Dobre, O., Hwang, W.J., 2022. Federated Learning for Smart Healthcare: A Survey. volume 55.

[88] Office for Civil Rights, U.D.o.H..H.S., 2003. OCR PRIVACY BRIEF SUMMARY OF THE HIPAA PRIVACY RULE HIPAA Compliance Assistance. Summary of HIPAA Privacy Rule URL: www.hhs.gov/sites/default/files/privacysummary.pdf.

[89] Oh, S.R., Seo, Y.D., Lee, E., Kim, Y.G., 2021. A comprehensive survey on security and privacy for electronic health data. International Journal of Environmental Research and Public Health 18.

[90] Omar, A.A., Bhuiyan, M.Z.A., Basu, A., Kiyomoto, S., Rahman, M.S., 2019. Privacy-friendly platform for healthcare data in cloud based on blockchain environment. Future Generation Computer Systems 95.

[91] Owens, K., Alem, A., Roesner, F., Kohno, T., 2022. Electronic Monitoring Smartphone Apps: An Analysis of Risks from Technical, Human-Centered, and Legal Perspectives. Proceedings of the 31st USENIX Security Symposium, Security 2022 .

[92] Packhäuser, K., Gündel, S., Münster, N., Syben, C., Christlein, V., Maier, A., 2022. Deep learning-based patient re-identification is able to exploit the biometric nature of medical chest X-ray data. Scientific Reports 12.

[93] Parikh, R.B., Teeple, S., Navathe, A.S., 2019. Addressing Bias in Artificial Intelligence in Health Care. JAMA 322.

[94] Park, J., Lee, D.H., 2018. Privacy preserving k-nearest neighbor for medical diagnosis in e-health cloud. Journal of Healthcare Engineering .

[95] Pew Charitable Trusts, 2016. Use of electronic offender-tracking devices expands sharply. Report available at: http://www.pewtrusts.org/en/research-and-analysis/issue-briefs/2016/09/use-of-electronic-offender-trackingdevices-expands-sharply .

[96] Premnath, S.N., Jana, S., Croft, J., Gowda, P.L., Clark, M., Kasera, S.K., Patwari, N., Krishnamurthy, S.V., 2013. Secret key extraction from wireless signal strength in real environments. IEEE Trans. on Mobile Computing 12.

[97] Price, W.N., 2018. Big data and black-box medical algorithms. Science Translational Medicine 10.

[98] Puppala, M., He, T., Chen, S., Ogunti, R., Yu, X., Li, F., Jackson, R., Wong, S.T.C., 2015. Meteor: An enterprise health informatics environment to support evidence-based medicine. IEEE Transactions on Biomedical Engineering 62.

[99] Puppala, M., He, T., Yu, X., Chen, S., Ogunti, R., Wong, S.T., 2016. Data security and privacy management in healthcare applications and clinical data warehouse environment. 3rd IEEE EMBS Int. Conf. on Biomedical and Health Informatics, BHI 2016 .

[100] Qiu, H., Qiu, M., Liu, M., Memmi, G., 2020. Secure Health Data Sharing for Medical Cyber-Physical Systems for the Healthcare 4.0. IEEE Journal of Biomedical and Health Informatics 24.

[101] Rajput, A.R., Li, Q., Taleby Ahvanooey, M., Masood, I., 2019. EACMS: Emergency Access Control Management System for Personal Health Record Based on Blockchain. IEEE Access 7.

[102] Ramirez, E., Brill, J., Ohlhausen, M.K., Wright, J., D McSweeny, T., 2014. DATA Brokers: A Call for TRansparency and Accountability. Federal Trade Commission .

[103] Rasmussen, K.B., Castelluccia, C., Heydt-Benjamin, T.S., Capkun, S., 2009. Proximity-based access control for implantable medical devices. Proceedings of the ACM Conference on Computer and Communications Security .

[104] Rathore, H., Mohamed, A., Al-Ali, A., Du, X., Guizani, M., 2017. A review of security challenges, attacks and resolutions for wireless medical devices. 2017 13th International Wireless Communications and Mobile Computing Conference, IWCMC 2017 .

[105] Reddy, S., Allan, S., Coghlan, S., Cooper, P., 2020. A governance model for the application of AI in health care. Journal of the American Medical Informatics Association 27.

[106] Reicin, E., 2022. Council post: Protecting consumer health data privacy beyond hipaa. URL: https://www.forbes.com/sites/forbesnonprofitcouncil/2022/05/10/protecting-consumer-health-data-privacy-beyond-hipaa/?sh=6927889e7b4e.

[107] Reveilhac, M., Blanchard, A., 2022. The framing of health technologies on social media by major actors: Prominent health issues and COVID-related public concerns. International Journal of Information Management Data Insights 2.

[108] Roan, D., 2016. Wiggins and froome medical records released by 'russian hackers'. URL: https://www.bbc.com/news/world-37369705.

[109] Saleheen, N., Ullah, M.A., Chakraborty, S., Ones, D.S., Srivastava, M., Kumar, S., 2021. WristPrint: Characterizing User Re-identification Risks from Wrist-worn Accelerometry Data. ACM Conference on Computer and Communications Security .

[110] Salem, O., Alsubhi, K., Shaafi, A., Gheryani, M., Mehaoua, A., Boutaba, R., 2022. Man-in-the-Middle Attack Mitigation in Internet of Medical Things. IEEE Transactions on Industrial Informatics 18.

[111] Sasson, E.B., Chiesa, A., Garman, C., Green, M., Miers, I., Tromer, E., Virza, M., 2014. Zerocash: Decentralized anonymous payments from bitcoin, in: 2014 IEEE symposium on security and privacy, IEEE.

[112] Schnell, R., Bachteler, T., Reiher, J., 2009. Privacy-preserving record linkage using Bloom filters. BMC Medical Informatics and Decision Making 9.

[113] Schnell, R., Borgs, C., 2020. Encoding hierarchical classification codes for privacy-preserving record linkage using bloom filters, in: Machine Learning and Knowledge Discovery in Databases. Springer International Publishing.

[114] Seh, A.H., Zarour, M., Alenezi, M., Sarkar, A.K., Agrawal, A., Kumar, R., Khan, R.A., 2020. Healthcare data breaches: Insights and implications. Healthcare (Switzerland) 8.

[115] Sharif, H.U., 2021. Privacy Preservation of Medical Data using Random Decision Tree 9.

[116] Sharma, P., Moparthi, N.R., Namasudra, S., Shanmuganathan, V., Hsu, C.H., 2021. Blockchain-based IoT architecture to secure healthcare system using identity-based encryption. Expert Systems .

[117] Sharrett, L., 2021. Kentucky mom alleges hospital workers missed her cancer — then covered up their mistake. URL: https://www.nbcnews.com/news/us-news/kentucky-mom-alleges-hospital-workers-missed-her-cancer-then-covered-n1258533.

[118] Shi, L., Yuan, J., Yu, S., Li, M., 2015. MASK-BAN: Movement-aided authenticated secret key extraction utilizing channel characteristics in body area networks. IEEE Internet of Things Journal 2.

[119] Sinhal, R., Ansari, I.A., 2023. Machine learning based multipurpose medical image watermarking. Neural Computing and Applications 5.

[120] Solove, B.D.J., 2011. Nothing to hide: The false tradeoff between privacy and security .

[121] Steinfeld, R., Bull, L., Zheng, Y., 2002a. Content extraction signatures .

[122] Steinfeld, R., Bull, L., Zheng, Y., 2002b. Content extraction signatures, in: Kim, K. (Ed.), Information Security and Cryptology — ICISC 2001, Springer Berlin Heidelberg, Berlin, Heidelberg. pp. 285–304.
Bose et al.: *Preprint submitted to Elsevier*    Page 24 of 26

A Survey on Privacy of Health Data Lifecycle

[123] Strydis, C., Seepers, R.M., Peris-Lopez, P., Siskos, D., Sourdis, I., 2013. A system architecture, processor, and communication protocol for secure implants. Transactions on Architecture and Code Optimization 10.

[124] Sudarsono, A., Yuliana, M., Darwito, H.A., 2017. A secure data sharing using identity-based encryption scheme for e-healthcare system. Proceeding - 2017 3rd International Conference on Science in Information Technology: Theory and Application of IT for Education, Industry and Society in Big Data Era, ICSITech 2017 2018-Janua.

[125] Sun, X., Zhang, P., Sookhak, M., Yu, J., Xie, W., 2017. Utilizing fully homomorphic encryption to implement secure medical computation in smart cities. Personal and Ubiquitous Computing 21.

[126] Sweeney, L., 2000. Simple demographics often identify people uniquely. Health (San Francisco) 671.

[127] Sweeney, L., 2002. A model for protecting privacy. Ieee S&P '02 10.

[128] Thomas, J., 2009. Medical records and issues in negligence. Indian Journal of Urology 25.

[129] Thummavet, P., Vasupongayya, S., 2013. A novel personal health record system for handling emergency situations. 2013 International Computer Science and Engineering Conference, ICSEC 2013 .

[130] Tian, H., He, J., Ding, Y., 2019. Medical Data Management on Blockchain with Privacy. Journal of Medical Systems 43.

[131] Vaiwsri, S., Ranbaduge, T., Christen, P., Schnell, R., 2022. Accurate privacy-preserving record linkage for databases with missing values. Information Systems 106.

[132] Vargas, L., Blue, L., Frost, V., Patton, C., Scaife, N., Butler, K.R.B., Traynor, P., 2019. Digital healthcare-associated infection: A case study on the security of a major multi-campus hospital system .

[133] Vora, J., Italiya, P., Tanwar, S., Tyagi, S., Kumar, N., Obaidat, M.S., Hsiao, K.F., 2018. Ensuring privacy and security in E-health records. CITS 2018 - 2018 International Conference on Computer, Information and Telecommunication Systems .

[134] Walker, K.L., 2016. Surrendering information through the looking glass: Transparency, trust, and protection. Journal of Public Policy and Marketing 35.

[135] Wan, Z., Hazel, J.W., Clayton, E.W., Vorobeychik, Y., Kantarcioglu, M., Malin, B.A., 2022. Sociotechnical safeguards for genomic data privacy. Nature Reviews Genetics 0123456789.

[136] Wang, W., Jiang, Y., Shen, Q., Huang, W., Chen, H., Wang, S., Wang, X., Tang, H., Chen, K., Lauter, K., Lin, D., 2019a. Toward scalable fully homomorphic encryption through light trusted computing assistance. arXiv:1905.07766.

[137] Wang, W., Shi, X., Qin, T., 2019b. Encryption-free Authentication and Integrity Protection in Body Area Networks through Physical Unclonable Functions. Smart Health 12.

[138] Wang, Z., Li, Q., Wang, Y., Liu, B., Zhang, J., Liu, Q., 2019c. Medical protocol security: DICOM vulnerability mining based on fuzzing technology .

[139] Wang, Z., Ma, P., Chi, Y., Zhang, J., 2018. Medical devices are at risk: Information security on diagnostic imaging system .

[140] Weisburd, K., Bhadha, V., Clauson, M., Elican, J., Kahn, F., Lawrenz, K., Pemberton, B., Ringler, R., Schaer, J., Sherman, M., 2021. Electronic Prisons: The Operation of Ankle Monitoring in the Criminal Legal System. GWU Legal Studies Research Paper URL: https://issuu.com/gwlawpubs/docs/electronic-prisons-report?fr=sOGI5NDcxODg3.

[141] Wolberg, W.H., Mangasarian, O.L., 1990. Multisurface method of pattern separation for medical diagnosis applied to breast cytology. Proceedings of the National Academy of Sciences 87.

[142] Wood, A., Najarian, K., Kahrobaei, D., 2020. Homomorphic Encryption for Machine Learning in Medicine and Bioinformatics. ACM Computing Surveys 53.

[143] Xie, Y., He, Q., Zhang, D., Hu, X., 2016. Medical ethics privacy protection based on combining distributed randomization with K-anonymity. Proceedings - 2015 8th International Congress on Image and Signal Processing, CISP 2015 .

[144] Xu, F., Qin, Z., Tan, C.C., Wang, B., Li, Q., 2011. IMDGuard: Securing implantable medical devices with the external wearable guardian. Proceedings - IEEE INFOCOM .

[145] Xu, J., Xue, K., Li, S., Tian, H., Hong, J., Hong, P., Yu, N., 2019. Healthchain: A Blockchain-Based Privacy Preserving Scheme for Large-Scale Health Data. IEEE Internet of Things Journal 6.

[146] Xue, L., Yu, Y., Li, Y., Au, M.H., Du, X., Yang, B., 2019. Efficient attribute-based encryption with attribute revocation for assured data deletion. Information Sciences 479.

[147] Yan, W., Hylamia, S., Voigt, T., Rohner, C., 2020. PHY-IDS: A physical-layer spoofing attack detection system for wearable devices. WearSys 2020 - Proceedings of the 6th ACM Workshop on Wearable Systems and Applications, Part of MobiSys 2020 .

[148] Yang, J.J., Li, J.Q., Niu, Y., 2015. A hybrid solution for privacy preserving medical data sharing in the cloud environment. Future Generation Computer Systems 43-44.

[149] Yang, Y., Xiao, X., Cai, X., Zhang, W., 2019. A secure and high visual-quality framework for medical images by contrast-enhancement reversible data hiding and homomorphic encryption. IEEE Access 7.

[150] Yao, A.C., 1982. Protocols for secure computations, in: 23rd annual symposium on foundations of computer science (sfcs 1982), IEEE.

[151] Yigzaw, K.Y., Olabarriaga, S.D., Michalas, A., Marco-Ruiz, L., Hillen, C., Verginadis, Y., de Oliveira, M.T., Krefting, D., Penzel, T., Bowden, J., Bellika, J.G., Chomutare, T., 2022. Health data security and privacy: Challenges and solutions for the future, in: Roadmap to Successful Digital Health Ecosystems. Elsevier.

[152] Yue, X., Wang, H., Jin, D., Li, M., Jiang, W., 2016. Healthcare Data Gateways: Found Healthcare Intelligence on Blockchain with Novel Privacy Risk Control. Journal of Medical Systems 40.

[153] Yuliana, M., Darwito, H.A., Sudarsono, A., Yofie, G., 2017. Privacy and security of sharing referral medical record for health care system. 2nd Int. Conf. on Science in Information Technology, ICSITech 2016: Information Science for Green Society and Environment .

[154] Yurdagul, M.A., Sencar, H.T., 2021. BLEKeeper: Response Time Behavior Based Man-In-The-Middle Attack Detection. Proceedings - 2021 IEEE Symposium on Security and Privacy Workshops, SPW 2021 .

[155] Zeadally, S., Isaac, J.T., Baig, Z., 2016. Security Attacks and Solutions in Electronic Health (E-health) Systems. Journal of Medical Systems 40.

[156] Zhang, A., Lin, X., 2018. Towards Secure and Privacy-Preserving Data Sharing in e-Health Systems via Consortium Blockchain. Journal of Medical Systems 42.

[157] Zhang, M., Chen, Y., Susilo, W., 2022. Decision Tree Evaluation on Sensitive Datasets for Secure E-Healthcare Systems. IEEE Transactions on Dependable and Secure Computing PP.

[158] Zhang, M., Raghunathan, A., Jha, N.K., 2013. MedMon: Securing medical devices through wireless monitoring and anomaly detection. IEEE Transactions on Biomedical Circuits and Systems 7.

[159] Zhang, Q., 2021. An overview and analysis of hybrid encryption: The combination of symmetric encryption and asymmetric encryption, in: 2021 2nd International Conference on Computing and Data Science (CDS).

[160] Zheng, Y., Hardjono, T., Pieprzyk, J., 1993. Sibling intractable function families and their applications, in: ReCALL. volume 1.